\documentstyle[10pt,epsf,epsfig,dp_delphititle]{dp_delphi}
\makeindex
\def\DpPaperGroup{EP}
\def\DpPaperRef{2000-027}
\def\DpDate{14 February 2000}
\def\DpAuthors{DELPHI Collaboration}
\def\DpSubmit{(Eur. Phys. J. C16(2000)229)}
\def\DpTitle{{A Study of the Lorentz Structure \\
 in Tau Decays}}
\def\DpComment{ }
\def\DpEMail{ }
\begin{document}
%%%%%%%%%%%%%%%%%%%%%%%%%% They are a problem with Coll.Sty ?
\makeatletter
%\input{dp_system:coll.sty}
% Collapse citation numbers to ranges.  Non-numeric and undefined labels
% are handled.  No sorting is done.  E.g., 1,3,2,3,4,5,foo,1,2,3,?,4,5
% gives 1,3,2-5,foo,1-3,?,4,5
\newcount\@tempcntc
\def\@citex[#1]#2{\if@filesw\immediate\write\@auxout{\string\citation{#2}}\fi
  \@tempcnta\z@\@tempcntb\m@ne\def\@citea{}\@cite{\@for\@citeb:=#2\do
    {\@ifundefined
       {b@\@citeb}{\@citeo\@tempcntb\m@ne\@citea\def\@citea{,}{\bf ?}\@warning
       {Citation `\@citeb' on page \thepage \space undefined}}%
    {\setbox\z@\hbox{\global\@tempcntc0\csname b@\@citeb\endcsname\relax}%
     \ifnum\@tempcntc=\z@ \@citeo\@tempcntb\m@ne
       \@citea\def\@citea{,}\hbox{\csname b@\@citeb\endcsname}%
     \else
      \advance\@tempcntb\@ne
      \ifnum\@tempcntb=\@tempcntc
      \else\advance\@tempcntb\m@ne\@citeo
      \@tempcnta\@tempcntc\@tempcntb\@tempcntc\fi\fi}}\@citeo}{#1}}
\def\@citeo{\ifnum\@tempcnta>\@tempcntb\else\@citea\def\@citea{,}%
  \ifnum\@tempcnta=\@tempcntb\the\@tempcnta\else
   {\advance\@tempcnta\@ne\ifnum\@tempcnta=\@tempcntb \else \def\@citea{--}\fi
    \advance\@tempcnta\m@ne\the\@tempcnta\@citea\the\@tempcntb}\fi\fi}
 
\makeatother
%%%%%%%%%%%%%%%%%%%%%%%%%% ??????????????????????????????????
% Generate the title page
\pagenumbering{roman}
\begin{titlepage}
\CERNpreprint{\DpPaperGroup}{\DpPaperRef} % Reference of the paper
\date{{\small\DpDate}} % Date of the paper
\title{\DpTitle} % Title of the paper
\address{\DpAuthors} % General name of the author(s)
\begin{shortabs} % Start the abstract
\noindent
%   abstract.tex
%
\noindent

This paper describes a measurement of the Michel parameters, $\eta$,
$\rho$, $\xi$, $\xi \delta$, and the average $\nu_{\tau}$
helicity, $h_{\nu_{\tau}}$, in $\tau$ lepton decays together with 
the first measurement of the tensor coupling in the weak charged current.
The $\tau^{+} \tau^{-}$ pairs were produced at the LEP $\mathrm{e^{+} e^{-}}$ 
collider at CERN from 1992 through 1995 in the DELPHI detector. 
Assuming lepton universality in the decays of the $\tau$ the measured values 
of the parameters were: 
$\eta =  -0.005 \pm 0.036 \pm 0.037$,
$\rho =  0.775 \pm 0.023 \pm 0.020$,
$\xi =  0.929 \pm 0.070 \pm 0.030$,
$\xi\delta = 0.779 \pm 0.070 \pm 0.028$,
$h_{\nu_{\tau}} =  -0.997 \pm 0.027 \pm 0.011$.
The strength of the tensor coupling was measured to be $\kappa^W_\tau = -0.029
\pm 0.036 \pm 0.018$. The first error is statistical and the second error
is systematic in all cases.
The results are consistent with the $V\!-\!A$ structure of the weak charged
current in decays of the $\tau$ lepton.

\end{shortabs}
\vfill
\begin{center}
\DpSubmit \ \\ % Horrible hack to allow to have DpSubmit empty
\DpComment \ \\
\DpEMail \ \\
\end{center}
\vfill
\clearpage
\headsep 10.0pt
\addtolength{\textheight}{10mm}
\addtolength{\footskip}{-5mm}
\begingroup
% Commands to process the author names
%
\newcommand{\DpName}[2]{\hbox{#1$^{\ref{#2}}$},\hfill}
\newcommand{\DpNameTwo}[3]{\hbox{#1$^{\ref{#2},\ref{#3}}$},\hfill}
\newcommand{\DpNameThree}[4]{\hbox{#1$^{\ref{#2},\ref{#3},\ref{#4}}$},\hfill}
\newskip\Bigfill \Bigfill = 0pt plus 1000fill
\newcommand{\DpNameLast}[2]{\hbox{#1$^{\ref{#2}}$}\hspace{\Bigfill}}
%
%\small
\footnotesize
\noindent
\DpName{P.Abreu}{LIP}
\DpName{W.Adam}{VIENNA}
\DpName{T.Adye}{RAL}
\DpName{P.Adzic}{DEMOKRITOS}
\DpName{Z.Albrecht}{KARLSRUHE}
\DpName{T.Alderweireld}{AIM}
\DpName{G.D.Alekseev}{JINR}
\DpName{R.Alemany}{VALENCIA}
\DpName{T.Allmendinger}{KARLSRUHE}
\DpName{P.P.Allport}{LIVERPOOL}
% ...
\normalsize
\endgroup
\titlefoot{Department of Physics and Astronomy, Iowa State
     University, Ames IA 50011-3160, USA
    \label{AMES}}
\titlefoot{Physics Department, Univ. Instelling Antwerpen,
     Universiteitsplein 1, B-2610 Antwerpen, Belgium \\
     \indent~~and IIHE, ULB-VUB,
     Pleinlaan 2, B-1050 Brussels, Belgium \\
     \indent~~and Facult\'e des Sciences,
     Univ. de l'Etat Mons, Av. Maistriau 19, B-7000 Mons, Belgium
    \label{AIM}}
\titlefoot{Physics Laboratory, University of Athens, Solonos Str.
     104, GR-10680 Athens, Greece
    \label{ATHENS}}
\titlefoot{Department of Physics, University of Bergen,
     All\'egaten 55, NO-5007 Bergen, Norway
    \label{BERGEN}}
\titlefoot{Dipartimento di Fisica, Universit\`a di Bologna and INFN,
     Via Irnerio 46, IT-40126 Bologna, Italy
    \label{BOLOGNA}}
\titlefoot{Centro Brasileiro de Pesquisas F\'{\i}sicas, rua Xavier Sigaud 150,
     BR-22290 Rio de Janeiro, Brazil \\
     \indent~~and Depto. de F\'{\i}sica, Pont. Univ. Cat\'olica,
     C.P. 38071 BR-22453 Rio de Janeiro, Brazil \\
     \indent~~and Inst. de F\'{\i}sica, Univ. Estadual do Rio de Janeiro,
     rua S\~{a}o Francisco Xavier 524, Rio de Janeiro, Brazil
    \label{BRASIL}}
\titlefoot{Comenius University, Faculty of Mathematics and Physics,
     Mlynska Dolina, SK-84215 Bratislava, Slovakia
    \label{BRATISLAVA}}
\titlefoot{Coll\`ege de France, Lab. de Physique Corpusculaire, IN2P3-CNRS,
     FR-75231 Paris Cedex 05, France
    \label{CDF}}
\titlefoot{CERN, CH-1211 Geneva 23, Switzerland
    \label{CERN}}
\titlefoot{Institut de Recherches Subatomiques, IN2P3 - CNRS/ULP - BP20,
     FR-67037 Strasbourg Cedex, France
    \label{CRN}}
\titlefoot{Now at DESY-Zeuthen, Platanenallee 6, D-15735 Zeuthen, Germany
    \label{DESY}}
\titlefoot{Institute of Nuclear Physics, N.C.S.R. Demokritos,
     P.O. Box 60228, GR-15310 Athens, Greece
    \label{DEMOKRITOS}}
\titlefoot{FZU, Inst. of Phys. of the C.A.S. High Energy Physics Division,
     Na Slovance 2, CZ-180 40, Praha 8, Czech Republic
    \label{FZU}}
\titlefoot{Dipartimento di Fisica, Universit\`a di Genova and INFN,
     Via Dodecaneso 33, IT-16146 Genova, Italy
    \label{GENOVA}}
\titlefoot{Institut des Sciences Nucl\'eaires, IN2P3-CNRS, Universit\'e
     de Grenoble 1, FR-38026 Grenoble Cedex, France
    \label{GRENOBLE}}
\titlefoot{Helsinki Institute of Physics, HIP,
     P.O. Box 9, FI-00014 Helsinki, Finland
    \label{HELSINKI}}
\titlefoot{Joint Institute for Nuclear Research, Dubna, Head Post
     Office, P.O. Box 79, RU-101 000 Moscow, Russian Federation
    \label{JINR}}
\titlefoot{Institut f\"ur Experimentelle Kernphysik,
     Universit\"at Karlsruhe, Postfach 6980, DE-76128 Karlsruhe,
     Germany
    \label{KARLSRUHE}}
\titlefoot{Institute of Nuclear Physics and University of Mining and Metalurgy,
     Ul. Kawiory 26a, PL-30055 Krakow, Poland
    \label{KRAKOW}}
\titlefoot{Universit\'e de Paris-Sud, Lab. de l'Acc\'el\'erateur
     Lin\'eaire, IN2P3-CNRS, B\^{a}t. 200, FR-91405 Orsay Cedex, France
    \label{LAL}}
\titlefoot{School of Physics and Chemistry, University of Lancaster,
     Lancaster LA1 4YB, UK
    \label{LANCASTER}}
\titlefoot{LIP, IST, FCUL - Av. Elias Garcia, 14-$1^{o}$,
     PT-1000 Lisboa Codex, Portugal
    \label{LIP}}
\titlefoot{Department of Physics, University of Liverpool, P.O.
     Box 147, Liverpool L69 3BX, UK
    \label{LIVERPOOL}}
\titlefoot{LPNHE, IN2P3-CNRS, Univ.~Paris VI et VII, Tour 33 (RdC),
     4 place Jussieu, FR-75252 Paris Cedex 05, France
    \label{LPNHE}}
\titlefoot{Department of Physics, University of Lund,
     S\"olvegatan 14, SE-223 63 Lund, Sweden
    \label{LUND}}
\titlefoot{Universit\'e Claude Bernard de Lyon, IPNL, IN2P3-CNRS,
     FR-69622 Villeurbanne Cedex, France
    \label{LYON}}
\titlefoot{Univ. d'Aix - Marseille II - CPP, IN2P3-CNRS,
     FR-13288 Marseille Cedex 09, France
    \label{MARSEILLE}}
\titlefoot{Dipartimento di Fisica, Universit\`a di Milano and INFN-MILANO,
     Via Celoria 16, IT-20133 Milan, Italy
    \label{MILANO}}
\titlefoot{Dipartimento di Fisica, Univ. di Milano-Bicocca and
     INFN-MILANO, Piazza delle Scienze 2, IT-20126 Milan, Italy
    \label{MILANO2}}
\titlefoot{Niels Bohr Institute, Blegdamsvej 17,
     DK-2100 Copenhagen {\O}, Denmark
    \label{NBI}}
\titlefoot{IPNP of MFF, Charles Univ., Areal MFF,
     V Holesovickach 2, CZ-180 00, Praha 8, Czech Republic
    \label{NC}}
\titlefoot{NIKHEF, Postbus 41882, NL-1009 DB
     Amsterdam, The Netherlands
    \label{NIKHEF}}
\titlefoot{National Technical University, Physics Department,
     Zografou Campus, GR-15773 Athens, Greece
    \label{NTU-ATHENS}}
\titlefoot{Physics Department, University of Oslo, Blindern,
     NO-1000 Oslo 3, Norway
    \label{OSLO}}
\titlefoot{Dpto. Fisica, Univ. Oviedo, Avda. Calvo Sotelo
     s/n, ES-33007 Oviedo, Spain
    \label{OVIEDO}}
\titlefoot{Department of Physics, University of Oxford,
     Keble Road, Oxford OX1 3RH, UK
    \label{OXFORD}}
\titlefoot{Dipartimento di Fisica, Universit\`a di Padova and
     INFN, Via Marzolo 8, IT-35131 Padua, Italy
    \label{PADOVA}}
\titlefoot{Rutherford Appleton Laboratory, Chilton, Didcot
     OX11 OQX, UK
    \label{RAL}}
\titlefoot{Dipartimento di Fisica, Universit\`a di Roma II and
     INFN, Tor Vergata, IT-00173 Rome, Italy
    \label{ROMA2}}
\titlefoot{Dipartimento di Fisica, Universit\`a di Roma III and
     INFN, Via della Vasca Navale 84, IT-00146 Rome, Italy
    \label{ROMA3}}
\titlefoot{DAPNIA/Service de Physique des Particules,
     CEA-Saclay, FR-91191 Gif-sur-Yvette Cedex, France
    \label{SACLAY}}
\titlefoot{Instituto de Fisica de Cantabria (CSIC-UC), Avda.
     los Castros s/n, ES-39006 Santander, Spain
    \label{SANTANDER}}
\titlefoot{Dipartimento di Fisica, Universit\`a degli Studi di Roma
     La Sapienza, Piazzale Aldo Moro 2, IT-00185 Rome, Italy
    \label{SAPIENZA}}
\titlefoot{Inst. for High Energy Physics, Serpukov
     P.O. Box 35, Protvino, (Moscow Region), Russian Federation
    \label{SERPUKHOV}}
\titlefoot{J. Stefan Institute, Jamova 39, SI-1000 Ljubljana, Slovenia
     and Laboratory for Astroparticle Physics,\\
     \indent~~Nova Gorica Polytechnic, Kostanjeviska 16a, SI-5000 Nova Gorica, Slovenia, \\
     \indent~~and Department of Physics, University of Ljubljana,
     SI-1000 Ljubljana, Slovenia
    \label{SLOVENIJA}}
\titlefoot{Fysikum, Stockholm University,
     Box 6730, SE-113 85 Stockholm, Sweden
    \label{STOCKHOLM}}
\titlefoot{Dipartimento di Fisica Sperimentale, Universit\`a di
     Torino and INFN, Via P. Giuria 1, IT-10125 Turin, Italy
    \label{TORINO}}
\titlefoot{Dipartimento di Fisica, Universit\`a di Trieste and
     INFN, Via A. Valerio 2, IT-34127 Trieste, Italy \\
     \indent~~and Istituto di Fisica, Universit\`a di Udine,
     IT-33100 Udine, Italy
    \label{TU}}
\titlefoot{Univ. Federal do Rio de Janeiro, C.P. 68528
     Cidade Univ., Ilha do Fund\~ao
     BR-21945-970 Rio de Janeiro, Brazil
    \label{UFRJ}}
\titlefoot{Department of Radiation Sciences, University of
     Uppsala, P.O. Box 535, SE-751 21 Uppsala, Sweden
    \label{UPPSALA}}
\titlefoot{IFIC, Valencia-CSIC, and D.F.A.M.N., U. de Valencia,
     Avda. Dr. Moliner 50, ES-46100 Burjassot (Valencia), Spain
    \label{VALENCIA}}
\titlefoot{Institut f\"ur Hochenergiephysik, \"Osterr. Akad.
     d. Wissensch., Nikolsdorfergasse 18, AT-1050 Vienna, Austria
    \label{VIENNA}}
\titlefoot{Inst. Nuclear Studies and University of Warsaw, Ul.
     Hoza 69, PL-00681 Warsaw, Poland
    \label{WARSZAWA}}
\titlefoot{Fachbereich Physik, University of Wuppertal, Postfach
     100 127, DE-42097 Wuppertal, Germany
    \label{WUPPERTAL}}
\addtolength{\textheight}{-10mm}
\addtolength{\footskip}{5mm}
\clearpage
\headsep 30.0pt
\end{titlepage}
%%%%%%%%%%%%%%%%%%%%%%%%%
%
% Change for the document body
%%\pagestyle{heading} % for page numbering
\pagenumbering{arabic} % page numbering in number
\setcounter{footnote}{0} %
\large
%\linenumbers %%%CD
%\linenumbers %%%CD
%   document.tex
%

\section{Introduction} \label{intro}

The Michel parameters~\cite{michel},~$\eta$,~$\rho$,~$\xi$, and $\xi
\delta$, are a set of experimentally accessible parameters which are
bilinear combinations of ten complex coupling constants describing the
couplings in the charged current decay of charged leptons. The
Standard Model makes a specific prediction about the exact nature of
the structure of the weak charged current.  $\tau$ leptons provide a
unique environment in which to verify this prediction. Not only is the
large mass of the $\tau$ lepton (and thus an extensive range of decay
channels) strong motivation to search for deviations from the Standard
Model but the $\tau$ also offers the possibility to test the
hypothesis of lepton universality.

The Michel parameters in $\tau$ decays have been extensively studied
by many experiments both at $\mathrm{e^+e^-}$ colliders running at the
$Z$ pole and at low energy
machines~\cite{pdg98a,opal}
%~\cite{argus1,argus2,argus3,argus4,argus5,argus6,cleo1,cleo2,cleo3,aleph,l31,l32,opal,sld}. 
This paper describes an
analysis of $\tau$ decays using both the purely leptonic and the
semi-leptonic (hadronic) decay modes, the latter being selected
without any attempt to identify the specific decay channel. By
grouping together all the semi-leptonic decays one can obtain a high
efficiency and purity at the expense of a loss of sensitivity to the
relevant parameters. This sensitivity is recuperated by splitting the
semi-leptonic decay candidates into bins of invariant mass of the
hadronic decay products, each bin being separately dominated by a
different $\tau$ decay mode.
Results are presented both with and without 
the assumption of lepton universality.

The measurement of the Michel parameters in the purely leptonic decay
modes of the $\tau$ allows limits to be placed on new physics. The
large number of Michel parameters, however, reduces the experimental
sensitivity in placing these limits. Moreover, the Michel
parameterisation does not cover the full variety of possible
interactions; in particular it does not include terms with
derivatives. However, a complementary test of a special type of new
interaction is presented. In addition to testing new couplings 
with leptonic currents that conserve fermion chiralities,
the possibility of an anomalous coupling of a leptonic charged tensor
current is explored.

%=============================

\boldmath
\section{The Michel parameters and $\nu_{\tau}$ helicity} 
\label{mich}
\unboldmath

The most general, lepton-number conserving,
derivative free, local, Lorentz invariant four-lepton 
interaction matrix element, $\cal{M}$, describing the leptonic decay 
$\tau \rightarrow l \overline{\nu}_l \nu_{\tau}$, ($l=$~e~or~$\mu$),  
can be written as follows~\cite{scheck,mursche,murroos}:

\begin{eqnarray}
\cal{M} & = &
\frac{4 G}{\sqrt{2}}
\sum_{N = S,V,T \atop{i,j = L,R}} g^N_{ij}
\left\langle    \overline{v}_{l_i} \left| \; \Gamma^N \; \right|    \left(
v_{\overline{\nu}_l}  \right)_n      \right\rangle
\left\langle ( \overline{v}_{\nu_\tau} )_m    \left| \; \Gamma_N \; \right|
u_{\tau_j}               \right\rangle  ,
\label{eqn:matele}
\end{eqnarray}
which is characterised by spinors of definite chirality. $G$ is a
coupling constant, 
%normally taken to be the Fermi coupling constant 
and the $\Gamma_N$ represent the various
forms of the weak charged current allowed by Lorentz invariance. 
The $n$ and $m$ in Eqn.~\ref{eqn:matele} are the chiralities of the
neutrinos which are uniquely determined by a given $N$, $i$ and $j$. In the
case of vector and axial-vector interactions the chirality of the neutrino is
equal to the chirality of its associated charged lepton, while it is the
opposite in the case of scalar, pseudoscalar and tensor interactions.
In all cases we refer to the helicities and chiralities of particles;
those of antiparticles are implicitly taken to have the opposite sign.

The $g^{N}_{ij}$ parameters are complex coupling constants. There
are 12 of these but, excluding the possibility of the existence of
a vector boson carrying a chiral charge, two of the constants,
$g^{T}_{LL}$ and $g^{T}_{RR}$, are identically zero. As the
couplings can be complex, with an arbitrary phase, there are 19
independent parameters. The Standard Model $V\!-\!A$ structure for the
weak charged current predicts that $g^{V}_{LL} = 1$ with all other
couplings being identically zero.
Neglecting phase space effects, the 
rate for the decay $\tau^- \to l^- \nu_\tau \bar{\nu}_l$ can be~written~\cite{fetscher,raab}~as
\begin{equation}
\label{eqn:decaywidth}
\Gamma(\tau^- \to l^- \nu_\tau \bar{\nu}_l) =  \frac {G^2 m_{\tau}^5} {192 \pi^3} \cdot \frac{A}{16},
\end{equation}
with the definition
\begin{eqnarray}
\label{eqn:normalisation}
 A   &  \equiv  &
    4 \bigr( |g^S_{RR}|^2 + |g^S_{LR}|^2 + |g^S_{RL}|^2 + |g^S_{LL}|^2 \bigl)
    + 48 \bigr( |g^T_{LR}|^2 + |g^T_{RL}|^2 \bigl)  \nonumber  \\
  & &  + 16 \bigr( |g^V_{RR}|^2 + |g^V_{LR}|^2 + |g^V_{RL}|^2 + |g^V_{LL}|^
2 \bigl)
 \hspace*{4mm} \equiv \hspace*{4mm} 16 .
\end{eqnarray}

From the above normalisation condition the maximum
values that the coupling constants $g^{N}_{ij}$ can take are 2, 1 and
1/$\sqrt{3}$ for $N=S,V$ and $T$ respectively.
The parameter $G$ has to be measured from
the decay rate and absorbs any deviation in the overall normalisation.
The shapes of the spectra and the ratios of branching ratios,
as used in this analysis, are insensitive to the 
overall normalisation and hence to $G$.

\begin{figure}[hbt]
\centering
  \epsfxsize=14.0cm
  \leavevmode
  \epsfbox[70 400 530 650]{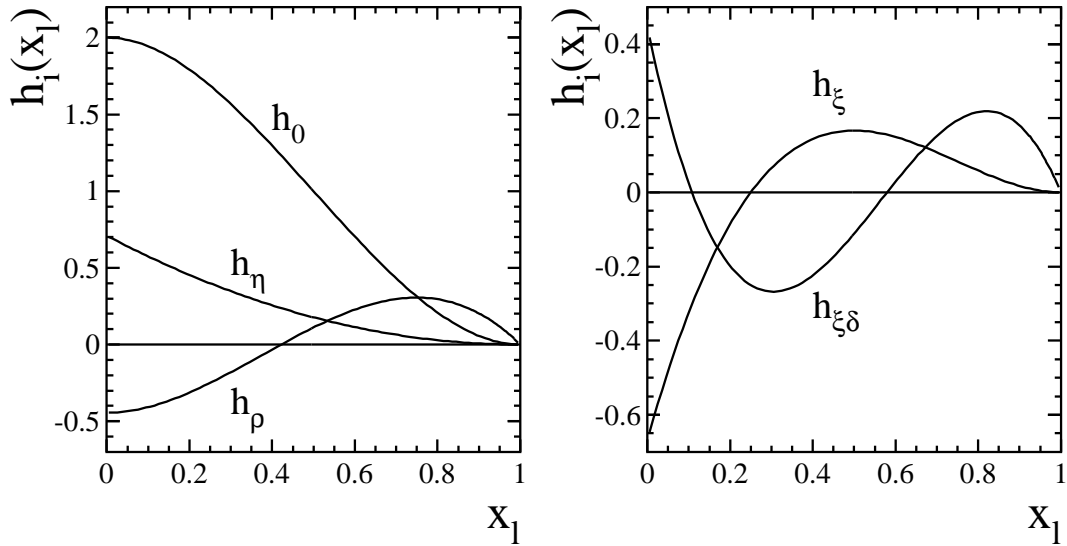}
%  \epsfbox[29 401 544 667]{poly_pap.ps}
 \caption{Polynomial functions in the laboratory frame for the $\tau \rightarrow
   l \overline{\nu}_{l} \nu_{\tau}$ decay channel at Born level.
   The $\eta$ polynomial is normalised to $m_{\mu} / m_{\tau}$.  } 
%   Effects of the
%   finite mass of the $\mu$ and radiative corrections are not included
%   in the plots. A description of how these effects are accounted for
%   in the fit is described later in the text.}
 \label{fig:polys}
\end{figure}

The matrix element written in Eqn.~\ref{eqn:matele} can be used to
form the decay distribution of the leptonic $\tau$ decay as follows:
%\begin{eqnarray}
%\label{eqn:leptdecay} \frac 1 \Gamma \frac {d \Gamma} {d x_l} & = &
%        H_0(x_l) -  {\cal P}_{\tau}  H_1(x_l) \\ 
%  & = & h_0(x_l) + \eta h_{\eta}(x_l)
%   + \rho h_{\rho}(x_l) \nonumber \\
%\label{eqn:leptdecay2} 
% &   &  \hspace{2cm} - {\cal P}_{\tau}  \left(
%\xi h_{\xi}(x_l) + \xi \delta h_{\xi \delta}(x_l) \right)
%\end{eqnarray}
\begin{eqnarray}
\label{eqn:leptdecay} 
\label{eqn:leptdecay2} 
      \frac 1 \Gamma \frac {d \Gamma} {d x_l} & = &
        H_0(x_l) -  {\cal P}_{\tau}  H_1(x_l) \\ 
  & = & [h_0(x_l) + \eta h_{\eta}(x_l)
   + \rho h_{\rho}(x_l) ]
    - {\cal P}_{\tau}  [
\xi h_{\xi}(x_l) + \xi \delta h_{\xi \delta}(x_l) ]
\end{eqnarray}
where $x_l = E_l / E_{max}$ is the normalised energy of the daughter
lepton and ${\cal P}_{\tau}$ is the average $\tau$ polarisation. $E_{max}$
is the maximum kinematically allowed energy of the lepton, $l$. In the 
rest frame of the $\tau$, $E_{max} = \frac { m_{\tau}^2 + m_l^2 } 
{ 2 m_{\tau} }$. In the laboratory frame $E_{max} \approx E_{\tau}$ or the beam 
energy $E_{beam}$.
The $h$'s at Born level are polynomials and are illustrated in
Fig.~\ref{fig:polys}. The Michel parameters, $\eta$, $\rho$, $\xi$ and
$\xi \delta$, are bilinear combinations of the complex coupling
constants \cite{michel} and take the following form 
in terms of the complex coupling constants:

\begin{eqnarray}
%%%\eta & = & \textstyle\frac{1}{2} \mbox{Re} \left(6 g^V_{RL} g^{T \ast}_{LR} + 6
%%%           g^V_{LR} g^{T \ast}_{RL} + g^S_{RR} g^{V \ast}_{LL} +
%%%           g^S_{RL} g^{V \ast}_{LR} + g^S_{LR} g^{V \ast}_{RL} +
%%%           g^S_{LL} g^{V \ast}_{RR}
%%%                           \right);    \\[1mm]
% eta corrected for revised version
\eta & = & \textstyle\frac{1}{2} \mbox{Re} \left(6 g^V_{LR} g^{T \ast}_{LR} + 6
           g^V_{RL} g^{T \ast}_{RL} + g^S_{RR} g^{V \ast}_{LL} +
           g^S_{RL} g^{V \ast}_{LR} + g^S_{LR} g^{V \ast}_{RL} +
           g^S_{LL} g^{V \ast}_{RR}                          \right);    \\[1mm]
%\rho & = & \textstyle\frac{3}{4} \left\{ 1 - |g^V_{RL}|^2 - |g^V_{LR}|^2
%           - 2|g^T_{RL}|^2 - 2|g^T_{LR}|^2
%          ~ -~ \mbox{Re} \bigr(g^S_{RL}g^{T\ast}_{RL} 
%           + g^S_{LR}g^{T\ast}_{LR} \bigl) \right\};    \\[1mm]
\rho & = & \textstyle\frac{3}{16} (4|g^V_{LL}|^2 + 4|g^V_{RR}|^2 + |g^S_{LL}|^2 + |g^S_{RR}|^2 
           +|g^S_{RL}-2g^T_{RL}|^2  +|g^S_{LR}-2g^T_{LR}|^2 );    \\[1mm]
\xi & = & -\textstyle\frac{1}{4} \bigr( |g^S_{RR}|^2 + |g^S_{LR}|^2 - |g^S_{RL}|^2 
              - |g^S_{LL}|^2 \bigl) ~+~ 5\bigr( |g^T_{LR}|^2 - |g^T_{RL}|^2 \bigl)
         \nonumber \\    
    &   &    - \bigr( |g^V_{RR}|^2 - 3|g^V_{LR}|^2 + 3|g^V_{RL}|^2 -
                |g^V_{LL}|^2 \bigl) ~+~ 4 
                  \mbox{Re}\bigr(g^S_{RL}g^{T\ast}_{RL} - g^S_{LR}g^{T\ast}_{LR} \bigl);  \\[2mm]
% old submitted   \mbox{Re}\bigr(g^S_{LR}g^{T\ast}_{LR} - g^S_{RL}g^{T\ast}_{RL} \bigl);  \\[2mm]
%orginal negative \mbox{Re}\bigr(g^S_{RL}g^{T\ast}_{RL} - g^S_{LR}g^{T\ast}_{LR} \bigl);  \\[2mm]
%\xi\delta & = & -\textstyle\frac{3}{16}\bigr( |g^S_{RR}|^2 -
%              |g^S_{RL}|^2
%            + |g^S_{LR}|^2 - |g^S_{LL}|^2 \bigl) ~-~ \textstyle\frac{3}{4}\bigr(
%              |g^V_{RR}|^2 - |g^V_{LL}|^2 \bigl)
%                \nonumber \\
%   &   &    + \textstyle\frac{3}{4}\bigr( |g^T_{RL}|^2 - |g^T_{LR}|^2\bigl) ~+~
%             \textstyle\frac{3}{4} 
%             \mbox{Re}\bigr(g^S_{LR}g^{T\ast}_{LR} - g^S_{RL}g^{T\ast}_{RL} \bigl) .
%%orginal negative \mbox{Re}\bigr( g^S_{RL}g^{T\ast}_{RL} -  g^S_{LR}g^{T\ast}_{LR} \bigl) .
\xi\delta & = & \textstyle\frac{3}{16} (4|g^V_{LL}|^2 - 4|g^V_{RR}|^2 + |g^S_{LL}|^2 - |g^S_{RR}|^2 
           +|g^S_{RL}-2g^T_{RL}|^2  - |g^S_{LR}-2g^T_{LR}|^2 ).
\end{eqnarray}
With the Standard Model predictions for these
coupling constants the Michel parameters $\eta$, $\rho$, $\xi$ and
$\xi \delta$ take on the values $0$, $\frac 3 4$, $1$ and $\frac 3 4$
respectively.

It is instructive to consider the physical significance of some of
these parameters. A single measurement of $\rho$ does not constrain
the form of the interaction. For example, if $\rho$ were to be
measured to be $\frac 3 4$, as is the case for the Standard Model
prediction, then this would not rule out any combination of the six
couplings $g^S_{LL}, \, g^S_{LR}, \, g^S_{RL}, \, g^S_{RR}, \,
g^V_{RR}, \, g^V_{LL}$ with the other couplings being zero. Indeed a
$V\!+\!A$ structure would have a value of $\rho$ of $\frac 3 4$. In this
case one must examine the other parameters.  For example, a $V\!+\!A$
structure would mean that the parameter $\xi$ would be equal to $-1$.
The values of the Michel parameters for several examples of
interaction types are given in Table~\ref{tab:int_types}.
\begin{table}[htb]
\begin{center}
\begin{tabular}{c|c|l|cccc}  \hline 
\multicolumn{2}{c|}{Vertices} & \multicolumn{1}{c}{Coupling} &
   \multicolumn{4}{|c}{Parameters} \\ \cline{1-2} \cline{4-7}
$\tau-\nu_{\tau}$ & $l-\nu_{l}$ & \multicolumn{1}{c}{Constants} &
  \multicolumn{1}{|c}{$\rho$} & \multicolumn{1}{c}{$\delta$}
  & \multicolumn{1}{c}{$\xi$} & \multicolumn{1}{c}{$\eta$}
  \\ \hline 
V-A  &  V-A  &  $g^V_{LL}$=1  &  $3/4$  & $3/4$ & 1
      & 0 \\
V  &  V  &  $g^V_{LL}$= $g^V_{RL}$=$g^V_{LR}$=$g^V_{RR}$=$1/2$
   &  $3/8$  & $3/4$ & 0  & 0   \\
A  &  A  &  $g^V_{LL}$= $-g^V_{RL}$=$-g^V_{LR}$=$g^V_{RR}$=$1/2$
   &  $3/8$  & $3/4$ & 0 & 0    \\
V+A  &  V+A  &  $g^V_{RR}$=1  &  $3/4$  & $3/4$ & -1
     & 0  \\
V  &  V-A  &  $g^V_{LL}$=$g^V_{LR}$=$1/ \sqrt{2}$  &
    $3/8$  & $3/16$ & 2   & 0   \\
A  &  V-A  &  $g^V_{LL}$=$-g^V_{LR}$=$1/ \sqrt{2}$  &
    $3/8$  & $3/16$ & 2  & 0    \\
V+A  &  V-A  &  $g^V_{LR}$=1  & 0  & 0  & 3 & 0    \\  \hline 
\end{tabular}
\caption{The couplings and
Michel parameter values for various mixtures of vector and axial-vector
coupling at the two vertices in the decay $\tau \rightarrow l \nu_{\tau}
\overline{\nu}_{l}$.}
\label{tab:int_types}
\end{center}
\end{table}                                     

The $\eta$ parameter is of particular interest. It is sensitive to the
low energy part of the decay lepton spectrum. It is practically
impossible to measure $\eta$ for $\tau \rightarrow \mathrm{e}
\nu_{\tau} \overline{\nu}_e$ decays because of a heavily suppressive
factor of $\frac {m_e} {m_{\tau}}$ in the $h_{\eta}$ polynomial. This
suppressive factor is of the order of $\simeq 1/17$ for $\tau
\rightarrow \mu \nu_{\tau} \overline{\nu}_{\mu}$ and hence all
sensitivity to $\eta$ is in this channel. The $h_\eta$ polynomial
receives contributions from the interference between {\it vector and
  scalar} and {\it vector and tensor} interactions and is therefore
particularly sensitive to non $V\!-\!A$ interactions. If $\eta \neq 0$
there would be two or more different couplings with opposite
chiralities for the charged leptons and this would result in
non-maximal parity and charge conjugation violation. In this case, if
$V\!-\!A$ is assumed to be dominant, then the second coupling could be a
Higgs type coupling with a right handed $\tau$ and muon~\cite{fetscher}.

The leptonic decay rates
of the $\tau$ lepton may be affected by the exchange of these
non-standard charged scalar particles~\cite{hollik} and these effects
can be conveniently expressed through the parameter
$\eta$~\cite{stahl,pich}. The generalised leptonic decay rate of the
$\tau$ becomes
\begin{eqnarray}
 \Gamma(\tau \rightarrow l \nu_{\tau} \overline{\nu}_{l}) =
 \frac{G_{l \tau}^{2}m_{\tau}^{5}}{192 \pi^3}\left[
f(\frac{m_l^2}{m_\tau^2})+
   4 \frac{m_l}{m_\tau} g(\frac{m_l^2}{m_\tau^2}) \eta \right]r_{RC}^{\tau}
\label{eqn:gammalept}
\end{eqnarray}
where $G_{l \tau}$ is the coupling of the $\tau$ to a lepton of type
$l$, and equals the Fermi coupling constant if lepton universality
holds. The functions $f$ and $g$ and the
quantity $r_{RC}^{\tau}$ are described in~\cite{pich}.
The parameter $r_{RC}^{\tau}$ is a factor due to electroweak radiative
corrections, which to a good approximation has the value $0.9960$ for
both leptonic decay modes of the $\tau$. 
The functions $f$ and $g$ are phase space factors.
The factor $f(\frac{m_{{l
      }}^2}{m_\tau^2})$ is equal to $1.0000$ for electrons and
$0.9726$ for muons. However, the function $4 \frac{m_{{\rm
      e}}}{m_\tau} g(\frac{m_{{\rm e}}^2}{m_\tau^2})$ equals $0.0012$,
whereas the value of $4 \frac{m_\mu}{m_\tau}
g(\frac{m_\mu^2}{m_\tau^2})$ is relatively large, equal to $0.2168$.
Hence, under the assumption of lepton universality, 
a stringent limit on $\eta$ in $\tau\to\mu\overline{\nu}_{\mu}
\nu_{\tau}$ decays can be set on the basis of the branching ratio
measurements, since to a good approximation (see discussion in 
section~\ref{sec:results}),
\begin{eqnarray}
\frac {Br ( \tau \rightarrow \mu \nu_{\tau} \overline{\nu}_{\mu} )}
      {Br ( \tau \rightarrow \mathrm{e} \nu_{\tau} \overline{\nu}_e )}
       & = &
f \biggl( \frac {m_{\mu}^2} {m_{\tau}^2} \biggr) + 4 \frac
{m_{\mu}} {m_{\tau}} g \biggl( \frac{m_{\mu}^2}{m_{\tau}^2}\biggr)
\eta_{\mu} .
\label{eqn:etabr}
\end{eqnarray}

The variable $P^{\tau}_R$ is defined as the probability that a
right handed $\tau$ will decay into a lepton of either handedness
\cite{fetscher}. This variable is related to the Michel~parameters
$\xi$ and $\xi \delta$ and to five of the complex coupling constants
in the following way:
\begin{eqnarray}
\hspace*{1cm} P^\tau_R \hspace*{3mm} & = &  \textstyle\frac{1}{4}|g^S_{RR}|^2 +
\textstyle\frac{1}{4}|g^S_{LR}|^2
  + |g^V_{RR}|^2 + |g^V_{LR}|^2 + 3|g^T_{LR}|^2  \nonumber \\
 & = &  \textstyle\frac{1}{2}[1 + \textstyle\frac{1}{3}\xi - \textstyle\frac{16}{9}\xi \delta ] .
\label{eqn:ptaur}
\end{eqnarray}
Hence the quantity $P^{\tau}_R$ is a measure of the contributions of
five coupling constants involving right handed $\tau$'s. One can
therefore see that measuring the parameters $\xi$ and $\xi \delta$ is
of considerable interest in studying the structure of the weak charged
currents.

The Michel parameters are restricted by
boundary conditions. The leptonic decay rate of the $\tau$ in
Eqn.~\ref{eqn:leptdecay} has to be positive definite. Certain
combinations of the Michel parameters lead to unphysical effects. It
has been shown~\cite{michconst,rouge1,bartoldus} that the following constraints must be
satisfied:
\begin{eqnarray}
0 \leq & \rho & \leq 1 , \label{eqn:ineq1} \\
|\xi| & \leq & 3 , \label{eqn:ineq2}\\
\rho - \left| \xi \delta \right| & \geq & 0  , \label{eqn:ineq3}\\
9 - 9\rho + \left| 7 \xi \delta  - 3 \xi \right| & \geq & 0 .\label{eqn:ineq4}
\end{eqnarray}                        
These inequalities describe the interior and surface of
a tetrahedron in ($\rho$,$\xi$,$\xi\delta$) space.
The first two conditions arise from the fact that the different
couplings in the definitions of the Michel parameters occur in
quadrature. The 3rd constraint can be found directly if the $\tau$ decay rate
in Eqn.~\ref{eqn:leptdecay2} is forced to be positive definite for all 
values of $x_l$. The 4th constraint is derived from the equations
of two of the surfaces of the physically allowed tetrahedron.
It is interesting to note that the Standard
Model values of the Michel parameters are consistently at the edge of
the allowed region (see Fig.~\ref{fig:contours}).  These relations
are used in Section~\ref{sec:interp} to place limits on the coupling
constants using the measured values of the Michel parameters.

The decay width of the semi-leptonic decays of the $\tau$ can be
written, assuming vector and axial-vector couplings at the decay
vertices~as
\begin{eqnarray}
\frac 1 \Gamma \frac {d \Gamma} {d x} & = &
H_0(x) +  {\cal P}_{\tau}  H_1(x) \nonumber \\
& = & H_0(x) - h_{\nu_{\tau}} {\cal P}_{\tau}  H_1(x)
\label{eqn:hadrdecay}
\end{eqnarray}
where $x$ is a polarisation sensitive variable in each decay channel.
For the case of $\tau \rightarrow \pi \nu_{\tau}$ this variable is
$\cos \theta^*$, the decay angle of the $\pi$ in the $\tau$ rest
frame, whilst for the two cases of $\tau \rightarrow \rho \nu_{\tau}$
and $\tau \rightarrow a_1 \nu_{\tau}$ the variable used is the
$\omega$ variable described in~\cite{rouge}.  The polarisation
parameter $h_{\nu_{\tau}}$ is defined as
\begin{equation}
h_{\nu_{\tau}} = \frac {2 {\rm Re} (v_{\tau} a_{\tau}^{*}) }
             { \left| v_{\tau} \right|^2 + \left| a_{\tau} \right|^2 }, 
\end{equation}
where $v_{\tau}$ and $a_{\tau}$ are the vector and axial-vector
couplings of the $\tau$ lepton to $W$ bosons. In the limit of a
massless $\nu_{\tau}$ this is equivalent to the $\tau$ neutrino
helicity.

Assuming that the boson exchanged in producing the $\tau^+
\tau^-$ pair only involves vector and axial-vector type couplings then
the helicities of the $\tau^+$ and $\tau^-$ are almost 100\%
anti-correlated. This fact is used to construct the correlated
spectra:
\begin{eqnarray}
\label{eqn:2d_dist}
\frac {1} {\Gamma} \frac {d^2 \Gamma} {d x_1 d x_2} & = &
 \frac {1 + {\cal P}_{\tau}} {2} \left( H_0(x_1) - H_1(x_1) \right)
               \left( H_0(x_2) - H_1(x_2) \right) + \nonumber \\
 & & \frac {1 - {\cal P}_{\tau}} {2} \left( H_0(x_1) + H_1(x_1) \right)
                                     \left( H_0(x_2) +  H_1(x_2) \right)
\end{eqnarray}
in terms of the polarisation sensitive variable $x$ (which is decay
channel dependent). For leptonic decays the $H_0$ and $H_1$ functions
are the polynomials described previously in Eqns.~\ref{eqn:leptdecay} and 
\ref{eqn:hadrdecay} and the 
polarisation sensitive variable is the scaled energy $x_l$.

It follows therefore that by performing two-dimensional fits over this
distribution one has full experimental access to all of the
Michel parameters together with the $\tau$ neutrino helicity,
$h_{\nu_{\tau}}$, and the $\tau$ polarisation, ${\cal P}_{\tau}$,
with the caveat that only vector and axial-vector currents are
assumed to contribute in the semileptonic decays.

\section{Anomalous tensor couplings}

The Lagrangian for the decay of the $\tau$ can be written in
the following way:
\begin{equation}
{\cal L} = \frac{g}{\sqrt{2}}~W^{\alpha}~
\left\{\overline{\tau} \gamma_{\alpha}
\frac{1-\gamma^{5}}{2}
\nu +
\frac{\kappa^W_\tau}{2m_{\tau}}
\partial_{\beta} \left(
\overline{\tau}\sigma_{\alpha\beta}
\frac{1-\gamma^{5}}{2}\nu \right) \right\} + {\mathrm h.c.}
\end{equation}
where $W^{\alpha}$ is the weak charged current of the decay
products of the W boson and $\kappa^{W}_{\tau}$ is a parameter which
controls the strength of the tensor coupling.  The choice of such a
kind of interaction to test for the existence of new physics is
inspired by experiments with semi-leptonic decays of pions \cite{bolo}
and kaons \cite{akim}, which show a deviation from the Standard Model
which can be explained by the existence of an anomalous interaction
with a tensor leptonic current \cite{pobla}. Since the new interaction
explicitly contains derivatives, its effect on the distortion of the
energy spectrum of charged leptons in $\tau$ decays can not be
described in terms of the known Michel parameters. Constraints will be
placed on the parameter $\kappa^W_\tau$ from the analysis of both
leptonic and semi-leptonic $\tau$ decays, fixing the Michel parameters
to their Standard Model values. The inclusion of the semi-leptonic
channels significantly increases the sensitivity to the new tensor
coupling and imposes stricter constraints.

For purely leptonic decays, the matrix element takes the form
\begin{eqnarray}
\cal{M} & = &
\frac{4 G}{\sqrt{2}}
\langle    \overline{v}_{l} \left| \; \gamma_\alpha \right| v_{\overline{\nu}_l} \rangle 
\left(    
\langle \overline{v}_{\nu_\tau} \left| \; \gamma_\alpha \right| u_{\tau_L} \rangle  
- i \frac{\kappa^W_\tau}{2m_\tau} q_\beta  
\langle \overline{v}_{\nu_\tau} \left| \; \sigma_{\alpha\beta} \right| u_{\tau_R} \rangle  
  \right)  ,
\label{eqn:mateletensor}
\end{eqnarray}
where $q$ is the momentum of the $W$.
The laboratory energy
spectrum of the charged decay~product can be expressed as
\begin{equation}
\frac{{\rm d}\Gamma}{{\rm d}x_l} \propto f(x_l) + {\cal P}_{\tau} g(x_l),
\label{eqn:width}
\end{equation}
where $x_l$ is again the normalised energy of the daughter lepton as 
defined in section~\ref{mich}. 

\begin{figure}[hbt]
\centering
  \leavevmode
  \epsfxsize=14.0cm
  \leavevmode
  \epsfbox[70 400 530 650]{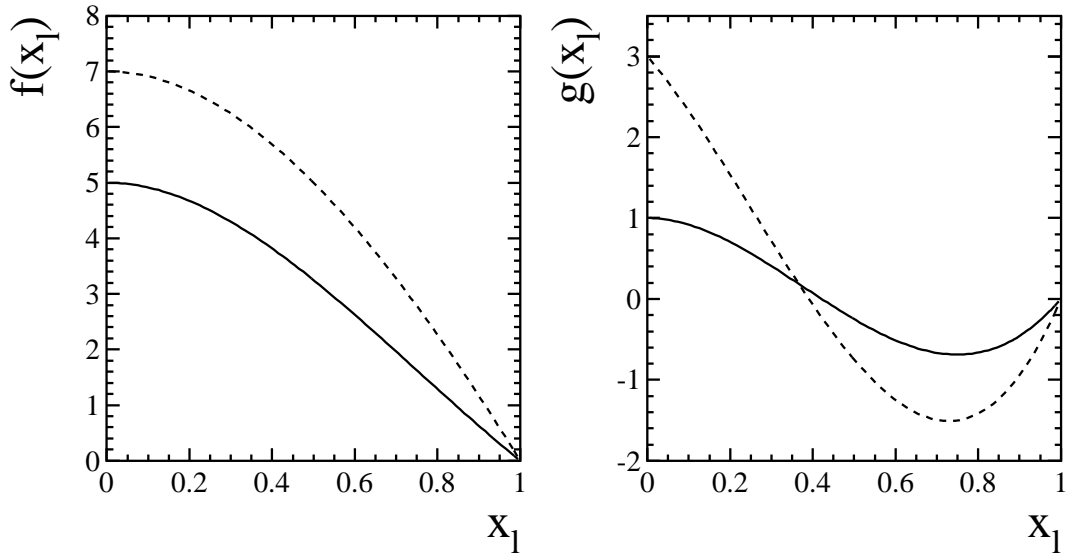}
 \caption{Polynomial functions  for the tensor coupling 
   contribution in the $\tau \rightarrow
   l \overline{\nu}_{l} \nu_{\tau}$ decay channel at Born level.
   Left plot shows $f(x_l)$ and right plot shows $g(x_l)$. In
   both plots the solid line illustrates the Standard Model case,
   $\kappa^W_\tau=0$, and the dashed line the case
   for~$\kappa^W_\tau=1$.   } 
 \label{fig:tensorfunc}
\end{figure}

The expressions for $f(x_l)$ and $g(x_l)$, accounting for the new tensor
interaction, were obtained in the rest frame of a decaying
lepton \cite{chiz}. Neglecting the mass of the final lepton and
boosting along the $\tau$ flight direction gives 
\begin{eqnarray}
f(x_l) & = & ~5 - 9x_l^2 + 4x_l^3 + 2\kappa^W_\tau (1-x_l^3), \nonumber \\
g(x_l) & = & ~1 -9x_l^2 + 8x_l^3 + 2\kappa^W_\tau (1 - 3x_l +2x_l^3).
\label{eqn:polys}
\end{eqnarray}
These functions are shown in
Fig.~\ref{fig:tensorfunc}.

In the Born
approximation the tensor interaction does not contribute to the
process $\tau \rightarrow \pi \nu_{\tau}$. 
Thus, among the main
$\tau$ semi-leptonic decay modes, only the decay modes 
$\tau \rightarrow \rho \nu_{\tau}
\rightarrow (2 \pi) \nu_{\tau}$ and $\tau \rightarrow a_1 \nu_{\tau}
\rightarrow (3 \pi) \nu_{\tau}$ yield information about the new tensor
interaction. The sensitivity can be increased by performing the analysis
using the two angular variables $\theta^*$ and $\psi$, 
where
$\theta^*$ is the angle between the emitted final (pseudo) vector particle after a Lorentz
transformation into the $\tau$ rest frame and the inverse of the three-vector component
of this Lorentz transformation.
%where
%$\theta^*$ is the angle between the $\tau$
%direction and the momentum of the final (pseudo) vector particle after a Lorentz
%transformation into the $\tau$ rest frame.
The angle $\psi$ is related to the angle of the $\rho$ or
$a_1$ decay products in the $\rho$ or $a_1$ system and is sensitive to
the polarisation of the hadronic system.  These two variables are
discussed in~\cite{rouge}.

The $\tau$ decay to a particle of spin 1 and mass $m_h$ and a neutrino
has two amplitudes, $A_L$ and $A_T$, representing longitudinal and
transverse polarisation of the spin-1 particle respectively.  From the
expression for the decay helicity amplitudes,
\begin{equation}
{\cal M}_\lambda \propto \overline{v}_{\overline{\nu}_l}(1+\gamma^{5})\left[\gamma_\alpha -
i\frac{\kappa^W_\tau}{2m_\tau}q_\beta\sigma_{\alpha\beta}\right]
u_{\tau}~\epsilon^*_\alpha(q,\lambda),
\end{equation}
where $\epsilon^*_\alpha$ is the polarisation vector of the spin 1
particle with momentum $q$ and helicity $\lambda$, one obtains
\begin{equation}
\frac{A_T}{A_L} = \frac{\sqrt{2}m_h}{m_\tau}~\frac{a_T}{a_L},
\end{equation}
where $a_T=1+\kappa^W_\tau/2$ and $a_L=1+(m^2_h/m^2_\tau) \kappa^W_\tau/2$.
Therefore for $\tau \rightarrow (2\pi)~\nu$ and $\tau \rightarrow (3\pi)~\nu$,
\begin{equation}
\frac{{\rm d}^2N}{{\rm d}\cos{\theta^*} ~ {\rm d}\cos{\psi}} \propto
(1+{\cal P}_\tau)H^+ + (1-{\cal P}_\tau)H^-,
\label{eqn:semilep}
\end{equation}
where
%\begin{eqnarray}
%H^+ & = & ~h_0(\psi)\left(a_L m_\tau \cos{\eta} \cos{\frac{\theta^*}{2}} +
%a_T m \sin{\eta} \sin{\frac{\theta^*}{2}}\right)^2 \nonumber \\ 
%  & + & ~h_1(\psi)\left[\left(a_L m_\tau \sin{\eta} \cos{\frac{\theta^*}{2}} -
%a_T m \cos{\eta} \sin{\frac{\theta^*}{2}}\right)^2 +
%a^2_T m^2 \sin^2\frac{\theta^*}{2}\right]  ,
%\end{eqnarray}
%\begin{eqnarray} 
%H^- & = & ~h_0(\psi)\left(a_L m_\tau \cos{\eta}
%\sin{\frac{\theta^*}{2}} -
%a_T m \sin{\eta} \cos{\frac{\theta^*}{2}}\right)^2 \nonumber \\ 
%  & + & ~h_1(\psi)\left[\left(a_L m_\tau \sin{\eta} \sin{\frac{\theta^*}{2}} +
%a_T m \cos{\eta} \cos{\frac{\theta^*}{2}}\right)^2 +
%a^2_T m^2 \cos^2\frac{\theta^*}{2}\right]  ,
%\end{eqnarray}
\begin{eqnarray}
H^+ & = & ~h_0(\psi)\left(a_L m_\tau \cos{\eta} \cos{\textstyle\frac{\theta^*}{2}} +
a_T m_h \sin{\eta} \sin{\textstyle\frac{\theta^*}{2}}\right)^2 \nonumber \\ 
  & + & ~h_1(\psi)\left[\left(a_L m_\tau \sin{\eta} \cos{\textstyle\frac{\theta^*}{2}} -
a_T m_h \cos{\eta} \sin{\textstyle\frac{\theta^*}{2}}\right)^2 +
a^2_T m^2_h \sin^2\textstyle\frac{\theta^*}{2}\right]  ,  \\[1mm]
H^- & = & ~h_0(\psi)\left(a_L m_\tau \cos{\eta}
\sin{\textstyle\frac{\theta^*}{2}} -
a_T m_h \sin{\eta} \cos{\textstyle\frac{\theta^*}{2}}\right)^2 \nonumber \\ 
  & + & ~h_1(\psi)\left[\left(a_L m_\tau \sin{\eta} \sin{\textstyle\frac{\theta^*}{2}} +
a_T m_h \cos{\eta} \cos{\textstyle\frac{\theta^*}{2}}\right)^2 +
a^2_T m^2_h \cos^2\textstyle\frac{\theta^*}{2}\right]  ,
\end{eqnarray}
and
\[ h_0(\psi) = \left\{ \begin{array}{l}
                         2\cos^2\psi \\
                         \sin^2\psi
                       \end{array}
               \right.  
                       \begin{array}[t]{l}
                         h_1(\psi) = \\
                       \end{array} 
               \left\{ \begin{array}{l}
                         2\sin^2\psi \\
                         (1+\cos^2\psi)/2
                       \end{array} 
               \right.  
                       \begin{array}[t]{l}
                         \mbox{for} \, \, \, \\
                       \end{array} 
               \left\{ \begin{array}{l}
                         \tau^- \rightarrow (2\pi)~\nu_{\tau} \\
                         \tau^- \rightarrow  (3\pi)~\nu_{\tau}
                       \end{array} 
               \right. .\]
Note that the angle $\eta$ used here
is unrelated to to the Michel parameter of the same name. 
Neglecting terms of ${\cal O}(m^2_\tau/E_\tau^2)$,  the relation between $\eta$ and
$\theta^*$ is
\begin{equation}
\cos{\eta} = \frac{m_\tau^2 - m^2_h + (m_\tau^2+m^2_h)\cos{\theta^*}}
{m_\tau^2 + m^2_h + (m_\tau^2-m^2_h)\cos{\theta^*}}.
\end{equation}

\section{The DELPHI Detector}
\label{sec:detector}
The DELPHI detector is described in detail elsewhere \cite{delphi,DELSIM}.
The following is a summary of the subdetector units particularly
relevant for this analysis.  All these covered the full solid angle of
the analysis except where specified.  In the DELPHI reference frame
the z-axis is taken along the direction of the e$^-$ beam. The angle
$\Theta$ is the polar angle defined with respect to the z-axis, $\phi$
is the azimuthal angle about this axis and $r$ is the distance from
this axis.  The reconstruction of a charged particle trajectory in the
barrel region of DELPHI resulted from a combination of the
measurements in:
\begin{itemize}
%\item the Vertex Detector (VD), made of three layers of silicon
%  micro-strip modules, at radii of 6.3, 9.0 and 11.0 cm from the beam
%  axis.  The space point precision was about $8~\mu$m in~r-$\phi$ and
%  varied from about 8~$\mu$m to 30~$\mu$m, depending on $\Theta$, in
%  r-z.  The two track resolution was 100~$\mu$m in r-$\phi$ and
%  200~$\mu$m in r-z.
% dwr update 16/12/99
\item the Vertex Detector (VD), made of three layers of silicon
  micro-strip modules, at radii of 6.3, 9.0 and 11.0 cm from the beam
  axis.  The space point precision in~r-$\phi$ was about $8~\mu$m, 
  while the  two track resolution was 100~$\mu$m.
  For the 1994 and 1995 data the innermost and outermost layers
  of the VD were equipped with double sided silicon modules,
  giving two additional measurements of the z coordinate.
\item the Inner Detector (ID), with an inner radius of 12 cm and an
  outer radius of 28 cm. A jet chamber measured 24 r-$\phi$
  coordinates and provided track reconstruction.  Its two track
  resolution in r-$\phi$ was 1~mm and its spatial precision $40~\mu$m.
  It was surrounded by an outer part which served mainly for
  triggering purposes. This outer part was replaced for the 1995 data
  with a straw-tube detector containing much less material.
\item the Time Projection Chamber (TPC), extending from 30~cm to
  122~cm in radius.  This was the main detector for the track
  reconstruction.  It provided up to 16 space points for pattern
  recognition and ionisation information extracted from 192 wires.
  Every $60^\circ$ in $\phi$ there was a boundary region between
  read-out sectors about $1^\circ$ wide which had no instrumentation.
  At cos$\Theta=0$ there was a cathode plane which caused a reduced
  tracking efficiency in the polar angle range
  $|$cos$\Theta|\!<\!0.035$. The TPC had a two track resolution of
  about 1.5~cm in r-$\phi$ and~in~z.  The measurement of the
  ionisation deposition had a typical precision of~$\pm 6\%$.
\item the Outer Detector (OD) with 5 layers of drift cells at a radius
  of 2~m from the beam axis, sandwiched between the RICH and HPC
  sub-detectors described below. Each layer provided a space point
  with 110~$\mu$m precision in r-$\phi$ and about 5~cm precision in~z.
\end{itemize}
These detectors were surrounded by a solenoidal magnet with a 1.2
Tesla field parallel to the z-axis. In addition to the detectors
mentioned above, the identification of the $\tau$ decay products
relied on:
\begin{itemize}
\item the barrel electromagnetic calorimeter, a High density
  Projection Chamber (HPC). This detector lay immediately outside the
  tracking detectors and inside the magnet coil. Eighteen radiation
  lengths deep for perpendicular incidence, its energy resolution was
  $\Delta E/E = 0.31/E^{0.44} \oplus 0.027$ where $E$ is in units of
  GeV.  It had a high granularity and provided a sampling of shower
  energies from nine layers in depth. It allowed a determination of
  the starting point of an electromagnetic shower with an accuracy of
  0.6 mrad in polar angle and 3.1 mrad in azimuthal angle.  The HPC
  had a modularity of $15^\circ$ in azimuthal angle. Between modules
  there was a region with a width of about~$1^\circ$ in azimuth where
  the energy resolution was degraded.  The HPC lay behind the OD and
  the Ring Imaging CHerenkov detector (RICH), not used in this analysis,
  which contained about 60\% of a radiation length.
\item the Hadron CALorimeter (HCAL), sensitive to hadronic showers and
  minimum ionising particles. It was segmented in 4 layers in depth,
  with a granularity of $3.75^{\circ}$ in polar angle and
  $2.96^{\circ}$ in azimuthal angle. Lying outside the magnet
  solenoid, it had a depth of 110 cm of iron.
\item the barrel muon chambers consisting of two layers of drift
  chambers, the first one situated after 90 cm of iron and the second
  outside the hadron calorimeter.  The acceptance in polar angle of
  the outer layer was slightly smaller than the other barrel detectors
  and covered the range $|$cos$\Theta|\!<\!0.602$.  The polar angle
  range $0.602\!<\!|$cos$\Theta|$ was covered by the forward muon
  chambers in certain azimuthal zones.
\end{itemize}
%
%Informaton  from
%the Ring-Imaging Cherenkov detector was not used in this analysis.
%Lying between  the TPC and OD in radius, it was 0.6 radiation
%lengths deep and 0.15 nuclear interaction lengths deep for
%particles of perpendicular incidence.
%It had an important effect on the performance of the
%calorimetry as it contained the majority of the material in the
%DELPHI barrel region inside the calorimeters.

The DELPHI trigger was very efficient for $\tau$ final states due to
the redundancy existing between its different components.  From the
comparison of the response of independent components, a trigger
efficiency of $(99.98\pm0.01)\%$ has been derived.

\section{Particle identification and energy calibration}

The detector response was extensively studied using simulated data
together with various test samples of real data where the identity of
the particles was unambiguously known. Examples of such samples
consisted of $\mathrm{e^+ e^- \rightarrow e^+ e^-}$, and $\mathrm{e^+
  e^-} \rightarrow \mu^+ \mu^-$ events together with the radiative
processes $\mathrm{e^+ e^- \rightarrow e^+ e^-} \gamma$ and
$\mathrm{e^+ e^-} \rightarrow \mu^+ \mu^- \gamma$. Test samples using
the redundancy of the detector were also used. An example of such a
sample is $\tau \rightarrow \pi (n \pi^0)$, $(n>0)$, selected by
tagging the $\pi^0$ decay in the HPC. This sample was extensively used
as a pure sample of charged hadrons to test the response of the
calorimetry and muon chambers.

\subsection{TPC ionisation measurement}

The ionisation loss of a track as it travels through the TPC gives
good separation between electrons and charged pions, particularly in
the low momentum range. Because of the importance of this variable it
was required that there were at least 28 anode wires used in the
measurement. This reduced the sample by a small amount primarily due
to particles being close to the boundary regions of the TPC sectors where
a narrow non-instrumented strip was located.  The $dE/dx$ pull
variable, $\prod^j_{dE/dx}$, for a particular particle hypothesis ($j$
= e, $\mu$, $\pi$, $K$) is defined as
\begin{equation}
{\textstyle \prod^{j}_{dE/dx}} = \frac { {dE/dx}_{meas} -
                                                  {dE/dx}_{expt}(j) }
                            { \sigma(dE/dx) }
\end{equation}
where ${dE/dx}_{meas}$ is the measured value, ${dE/dx}_{expt}(j)$ is
the expected momentum dependent value for a hypothesis $j$ and
$\sigma(dE/dx)$ is the resolution of the measurement.

\subsection{Electromagnetic calorimetry}

The HPC is used for $\mathrm{e},\gamma$
and $\pi^0$ identification. For charged particles $E_{ass}$ is the
energy deposited in the HPC. For electrons this energy
should be (within experimental errors) equal to the measured value of
the momentum.  Muons, being minimum ionising particles,
deposit only a small amount of energy in the calorimeter.
Most charged hadrons interact deep in the HPC or
in the HCAL and thus look like a minimum ionising particle in the
early part or all of the HPC, with an increased energy 
deposition in the later layers if an interaction occurs in the HPC.

The ratio of the energy deposition in the HPC to the reconstructed
momentum has a peak at one for electrons and a rising distribution
towards zero for hadrons. The pull variable, ${\textstyle
  \prod_{E/p}}$, is defined as
\begin{equation}
{\textstyle \prod_{E/p}} = \frac {E_{ass}/p^{\prime} - 1}
                            {\sigma(E_{ass}/p^{\prime};E_{ass})}
\end{equation}
where $p^{\prime}$ is the momentum refit without the use of the OD,
described in Section~\ref{sec:mom} below, and
$\sigma(E_{ass}/p^{\prime};E_{ass})$ is the expected resolution 
on $E_{ass}/p^{\prime}$ for an
electron with associated energy $E_{ass}$. 
This variable gives particularly
good separation at high momenta.

\subsection{Hadron calorimetry and muon identification}

\begin{figure}[hbt]
\centering
  \epsfxsize=14.0cm
    \leavevmode
      \epsfbox[50 400 540 665]{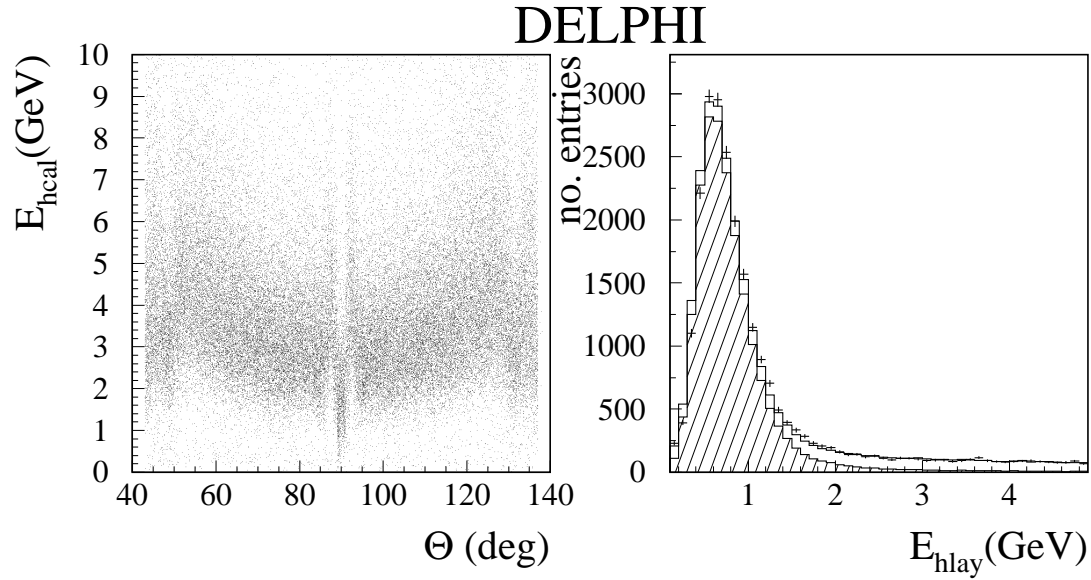}
\caption{The HCAL response to muons
  (left plot) together with the variable $E_{hlay}$ (right plot) for a
  sample of hadrons and muons in 1994 data (barrel region only).  The
  crosses are the data, the solid histogram is the simulated sum
  of hadrons and muons and the hatched area is the simulated muons.}
 \label{fig:ehlay}
\end{figure}

The HCAL was used in particular for separating pions from muons. As
muons travel through the HCAL they deposit a small amount of energy
evenly through the 4 layers and travel on into the muon chambers
whereas hadrons deposit all their energy late in the HPC and/or in the
first layers of the HCAL so that they rarely penetrate through to the
muon chambers.  Therefore muons can be separated from hadrons by
demanding energy associated to the particle in the last layer of the
HCAL together with an associated hit in the muon chambers. To further
distinguish muons from hadrons one can construct the variable
$E_{hlay}$, the average energy deposited in the HCAL per HCAL layer
defined as:
\begin{equation}
E_{hlay} = \frac {E_{HCAL}} {N_{layers}} \times \sin^2 \Theta
\end{equation}
where $E_{HCAL}$ is the total deposited energy in the HCAL;
$N_{layers}$ is the number of HCAL layers with an energy deposit and
$\sin^2 \Theta$ smoothes out the angular dependence of the energy
response of the HCAL (see Fig.~\ref{fig:ehlay}).  This variable can be
seen in Fig.~\ref{fig:ehlay}. Note that the step behaviour
around polar angles of $50^\circ$ and $130^\circ$ is due to
the reduction in the number of layers hit in the HCAL where
a muon passes through a mixture of barrel 
geometry and end-cap geometry. 

\subsection{Momentum determination and scale}
\label{sec:mom}

A good knowledge of the momentum and energy of charged particles is
required for a Michel parameter analysis. This is especially true for
the leptonic channels. As already mentioned the momentum is measured
by tracking the particles in a magnetic field as they traverse the
detector.  The precision on the component of momentum transverse to
the beam direction, $p_t$, obtained with the DELPHI tracking detectors
was $\Delta (1/p_t) = 0.0008 \rm{(GeV/c)^{-1}}$ for particles (except
electrons) with the same momentum as the beam. Calibration of the
momentum is performed with $\mathrm{e^+ e^-} \rightarrow \mu^+ \mu^-$
events. For lower momenta the masses of the $K^0_s$ and the $J/ \psi$
are reconstructed to give an absolute momentum scale for particles
other than electrons estimated, to a precision of 0.2\% over the full
momentum range.

The determination of the momentum of electrons is more
complicated. In passing through the RICH from the TPC to the OD,
particles traverse about 60\% of a radiation length. A large fraction
of electrons therefore lose a substantial amount of energy through
bremsstrahlung before they reach the OD.  Due to this the standard
momentum measurement of electrons would always tend to be biased to
lower values. This effect is somewhat reduced through only using the
measured momentum without using the OD, $p^{\prime}$. The result is
that this ``refit momentum'' shows a more Gaussian behaviour than the
standard momentum fit. The best estimate for the momentum of the
electron, $p_{el}$, is constructed in such a way as to benefit from
the better resolution of the momentum measurement at low momentum and
the smaller bremsstrahlung bias of the electromagnetic energy
measurement. The reconstructed momentum and the electromagnetic
energy were combined through a weighted average which took into
account the downward biases of the two respective measurements. The
energy of the radiated photons was also added to the electromagnetic
energy measurement to reduce further the effects of bremsstrahlung.

An algorithm was used which performed a weighted average depending on
the value of $E_{ass}/p^{\prime}$. The further this value was from
unity, the more the weight of the estimator with the lower value was
down scaled relative to the other. The scaling factor was inversely
proportional to the square of the number of standard deviations by
which the value of $E_{ass}/p^{\prime}$ differed from unity.

Subsequent references to the momenta of electrons imply the use of the
best estimator $p_{el}$. The momenta of other particles are measured
using the standard momentum fit, $p$, of the particle as it traverses
the detector.

\section{The selection of the event sample}

In order to determine the Michel parameters, a sample of exclusively
selected
leptonic decays of the $\tau$ together with an inclusive sample of
semi-leptonic decays have been used. The data sample corresponds to the
data
taken by DELPHI during 1992 (22.9~$pb^{-1}$ at $E_{cm}=$~91.3~GeV), 1993
(15.7~$pb^{-1}$ at $E_{cm}=$~91.2~GeV, 9.4~$pb^{-1}$ at $E_{cm}=$~89.2~GeV
and 4.5~$pb^{-1}$ at $E_{cm}=$~93.2~GeV), 1994 (47.4~$pb^{-1}$ at
$E_{cm}=$~91.2~GeV) and 1995 (14.3~$pb^{-1}$ at $E_{cm}=$~91.2~GeV,
9.2~$pb^{-1}$ at $E_{cm}=$~89.2~GeV and 9.3~$pb^{-1}$ at
$E_{cm}=$~93.2~GeV).

In all analyses, samples of simulated events were used which had been
passed
through a detailed simulation of the detector response~\cite{DELSIM} and
reconstructed with the same program as the real data. The Monte Carlo
event generators used were: KORALZ~4.0~\cite{KORALZ} together 
with the TAUOLA~2.5~\cite{tauola} $\tau$ decay package
for $\mathrm{e^+ e^-} \rightarrow \tau^+ \tau^-$ events; DYMU3~\cite{DYMU3} 
for $\mathrm{e^+ e^-} \rightarrow \mu^+ \mu^-$ events; BABAMC~\cite{BABAMC} 
for $\mathrm{e^+ e^- \rightarrow e^+ e^-}$ events;
JETSET 7.3~\cite{JETSET} for $\mathrm{e^+ e^-} \rightarrow q \overline{q}$ 
events; Berends-Daverveldt-Kleiss~\cite{BDK} for 
$\mathrm{e^+ e^- \rightarrow e^+ e^- e^+e^-}$, 
$\mathrm{e^+ e^-\rightarrow e^+ e^-} \mu^+ \mu^-$ and 
$\mathrm{e^+ e^- \rightarrow e^+ e^-} \tau^+ \tau^-$ events;
TWOGAM~\cite{teddy} for $\mathrm{e^+ e^- \rightarrow e^+ e^- q \bar{q}}$ events.

 The variables used in the initial preselection of the $\tau$ sample together with
the selection of the various decay channels are described
below.

\boldmath
\subsection{The $\mathrm{e^+ e^-} \rightarrow \tau^+ \tau^-$ sample}
\unboldmath
At LEP energies, a $\tau^+ \tau^-$ event appears as two highly
collimated low multiplicity jets in approximately opposite directions.
An event was separated into hemispheres by a plane perpendicular to
the event thrust axis, where the thrust was calculated using all
charged particles with momentum greater than 0.6 GeV/c. To be included
in the sample, it was required that the highest momentum charged
particle in at least one of the two hemispheres lie in the polar angle
range $|\cos \Theta|<0.732$.

Background from $\mathrm{e^+ e^-} \rightarrow q \overline{q}$ events
was reduced by requiring a charged particle multiplicity less than six
and a minimum thrust value of 0.996. The $\mathrm{e^+ e^-} \rightarrow
q \overline{q}$ background is however negligible in the
analysis of the Michel parameters as one is looking for events with
only one charged particle in each hemisphere.

Cosmic rays and beam gas interactions were rejected by requiring that
the highest momentum charged particle in each hemisphere have a point
of closest approach to the interaction region less than 4.5~cm in $z$
and less than 1.5~cm in the $r-\phi$ plane. It was furthermore
required that these particles have a difference in $z$ of their points
of closest approach at the interaction region of less than 3~cm. The
offset in $z$ of tracks in opposite hemispheres of the TPC was
sensitive to the time of passage of a cosmic ray event with respect to
the interaction time of the beams. The background left in the selected
sample was computed from the data by interpolating the distributions
outside the selected regions.

Two-photon events were removed by requiring a total energy in the
event, $E_{vis}$, greater than 8 GeV and a total transverse component of the
vector sum of the charged particle momenta in the event, $p^{miss}_t$, 
greater than 0.4 GeV/c.

Contamination from $\mathrm{e^+ e^- \rightarrow e^+ e^-}$ and
$\mathrm{e^+ e^-} \rightarrow \mu^+ \mu^-$ events was reduced by
requiring that the event acollinearity, $\theta_{acol} = \cos^{-1} ( -
\frac {p_1 \cdot p_2} {|p_1| |p_2|} )$, be greater than $0.5^\circ$.
The variables $p_1$ and $p_2$ are the momenta of the highest momenta
charged particles in hemisphere 1 and 2 respectively.

The $\mathrm{e^+ e^- \rightarrow e^+ e^-}$ background is reduced in
the second instance with a cut on the radial energy $E_{rad}$ (defined
as $E_{rad}=\sqrt{{E_1}^2+{E_2}^2} / E_{beam}$ where $E_1$ and $E_2$
are the energies deposited in the HPC in a $30^\circ$ cone around the
highest momentum charged particle in each hemisphere and $E_{beam}$ is
the beam energy). Events are retained if $E_{rad}<1$.

The $\mathrm{e^+ e^-} \rightarrow \mu^+ \mu^-$ background is reduced
in the second instance with a cut on the radial momentum $p_{rad}$
(defined as $p_{rad}=\sqrt{{p_1}^2+{p_2}^2} / p_{beam}$ where $p_1$
and $p_2$ are the momenta of the highest momentum charged particles in
each hemisphere and $p_{beam}$ is the beam momentum). Cutting on this
quantity is also effective in reducing the $\mathrm{e^+ e^-
  \rightarrow e^+ e^-}$ background. Events are retained if
$p_{rad}<1$.

As a result of the above selection $\sim 93000$ $\mathrm{e^+ e^-}
\rightarrow \tau^+ \tau^-$ candidates were selected from the 1992 to
1995 data set. The efficiency of selection in the $4\pi$ solid angle
was $\sim 54\%$. The background arising from $\mathrm{e^+ e^-
  \rightarrow e^+ e^-}$ events was estimated to be $(1.07\pm0.32)\%$,
from $\mathrm{e^+ e^-} \rightarrow \mu^+ \mu^-$ events
$(0.30\pm0.09)\%$ and from four-fermion processes $(0.93\pm0.28)\%$.
The background from $\mathrm{e^+ e^-} \rightarrow \mathrm{e^+ e^- q \bar{q}}$
was negligible. 
Since the efficiencies and backgrounds varied slightly from year to year
the data sets were treated independently.

\boldmath
\subsection{The $\tau \rightarrow \mathrm{e} \overline{\nu}_e \nu_{\tau}$ 
channel}
\unboldmath
The $\tau \rightarrow \mathrm{e} \overline{\nu}_e \nu_{\tau}$ decay
has the signature of an isolated charged particle which produces an
electromagnetic shower in the calorimetry. The produced electrons are
ultra-relativistic and leave an ionisation deposition in the Time
Projection Chamber corresponding to the plateau region above the
relativistic rise. Backgrounds from other $\tau$ decays arise
principally from one-prong hadronic decays where either the hadron
interacts early in the electromagnetic calorimeter or an accompanying
$\pi^{0}$ decay is wrongly associated to the charged particle track.

As an initial step in electron identification
it was required that there be one charged
particle in the hemisphere with a momentum greater than
$0.01p_{beam}$. To ensure optimal use of the HPC it was required that
the track lie in the polar angle range $0.035 < \left| \cos{\Theta}
\right| < 0.707$ and that the track extrapolation to the HPC should
lie outside any HPC azimuthal boundary region, as described in
Section~\ref{sec:detector}.

The $dE/dx$ measurement is crucial to the analysis and so it was required
that there were at least 28 anode wires with ionisation information in the
TPC. It was required that the $dE/dx$ measurement be consistent with that
of an electron by requiring that the ${\prod^e}_{dE/dx}$ variable be greater
than -2. This requirement was efficient, especially at low momentum, in
retaining signal and removing backgrounds from muons and hadrons.

\begin{figure}[hbt]
\centering
  \epsfxsize=14.0cm
  \leavevmode
  \epsfbox[37 404 560 670]{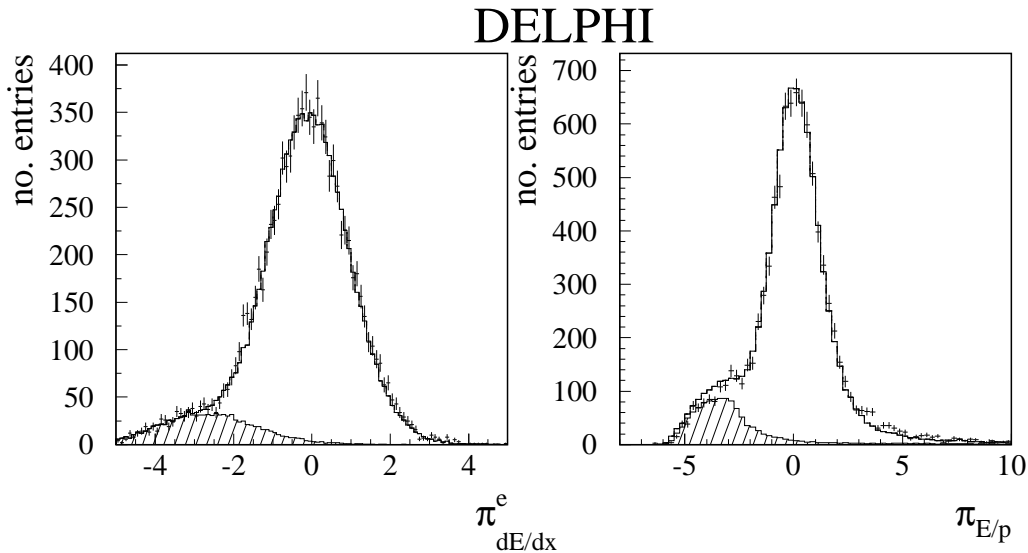}
 \caption{The ${\prod^e}_{dE/dx}$ and ${\prod}_{E/p}$ variables after
   application of all the other selection cuts except the one shown
   for 1994 data. The crosses are the data, the solid histogram is
   the sum of the signal and background and the shaded area is the
   background from $\tau \not\rightarrow \mathrm{e} \nu
   \overline{\nu}$ events.}
 \label{fig:vars_ele}
\end{figure}

The selection continued with a logical ``OR'' of two
criteria, the first on the ${\prod^{\pi}}_{dE/dx}$ variable, which
was particularly good at low momentum, and the second on the
${\prod}_{E/p}$ variable, which was particularly good at high
momentum. A particle was taken to be an electron if it deposited
greater than 0.5 GeV in the HPC and the value of ${\prod}_{E/p}$ was
greater than -2 ``OR'' the measured value of ${\prod^{\pi}}_{dE/dx}$
was greater than 3 and the momentum was greater than 0.01$p_{beam}$.
The ``OR'' thus gives a high constant efficiency over the whole
momentum range. The ${\prod^e}_{dE/dx}$ and ${\prod}_{E/p}$ variables
can be seen in Fig.~\ref{fig:vars_ele}.

The remaining background was reduced by requiring that there
be no hits in the muon chambers and no deposited energy beyond the
first HCAL layer. Residual background from $\tau \rightarrow \pi (n
\pi^0)\nu$ was reduced by cutting on the energy of the most energetic
neutral shower in the HPC observed in an $18^\circ$ cone around the
track. Neutral showers were not included in this requirement if they
were within $1^\circ$ of the track and hence compatible with being
bremsstrahlung photons.

The identification criteria were studied on test samples of real data.
The efficiency of the $dE/dx$ and HPC cuts were tested across the
whole momentum range by exploiting the redundancy of the two. Since
the simulation showed that the two measurements were instrumentally
uncorrelated, the overall bin by bin efficiency was calculated from
these two independent measurements.

Backgrounds arising from non-$\tau$ sources consisted of
$\mathrm{e^+e^- \rightarrow e^+e^-}$ and four-fermion $\mathrm{e^+e^-
  \rightarrow e^+e^- e^+e^-}$ events. The $\mathrm{e^+e^- \rightarrow
  e^+e^-}$ background was suppressed by the standard $\tau$
preselection cuts, {\it i.e.}  $p_{rad}<1$ and $E_{rad}<1$.
Four-fermion events remaining after the $E_{vis}$ and $P_{t}^{miss}$
cuts were further suppressed by demanding that if the $\tau
\rightarrow \mathrm{e} \nu_\tau \bar{\nu}_{\mathrm{e}}$ candidate had a 
momentum less than $0.2E_{beam}$ and there was only one particle detected 
in the opposite hemisphere with similarly a momentum below
$0.2E_{beam}$ then the $\tau \rightarrow \mathrm{e} \nu_\tau 
\bar{\nu}_{\mathrm{e}}$ candidate was retained if $\prod^{\pi}_{dE/dx}$ 
for the particle in the opposite hemisphere was less than 3 and therefore 
inconsistent with being an electron.

Application of the above procedure on the 1992 to 1995 data
resulted in a sample of $\sim 21500$ $\tau \rightarrow
\mathrm{e} \overline{\nu}_e \nu_{\tau}$ candidates. 
The efficiency of selection within the $4\pi$
angular acceptance was $35\%$.  The background arising from $\tau
\not\rightarrow \mathrm{e} \overline{\nu}_e \nu_{\tau}$ processes was
estimated to be $(3.89\pm1.17)\%$, from $\mathrm{e^+e^- \rightarrow
  e^+e^-}$ events $(1.61\pm0.48)\%$ and from $\mathrm{e^+e^-
  \rightarrow e^+e^- e^+e^-}$ events $(0.53\pm0.16)\%$.

\boldmath
\subsection{The $\tau \rightarrow \mu \overline{\nu}_{\mu} \nu_{\tau}$ channel}
\unboldmath
A muon candidate in the decay $\tau \rightarrow \mu
\overline{\nu}_{\mu} \nu_{\tau}$ appears as a minimum ionising
particle in the hadron calorimeter, penetrating through to the muon
chambers. Due to ionisation loss, a minimum momentum of about 2 GeV/c
is required for a muon to pass through the hadron calorimeter and into
the muon chambers.
It was therefore required that there be one charged particle
in the hemisphere with sufficient energy to penetrate through the
detector into the muon chambers. The candidate had to have a
momentum greater than $0.05p_{beam}$ and lie within the polar angle
interval $0.035 < \left| \cos{\Theta} \right| < 0.732$.

\begin{figure}[hbt]
\centering
  \epsfxsize=14.0cm
  \leavevmode
  \epsfbox[37 404 560 670]{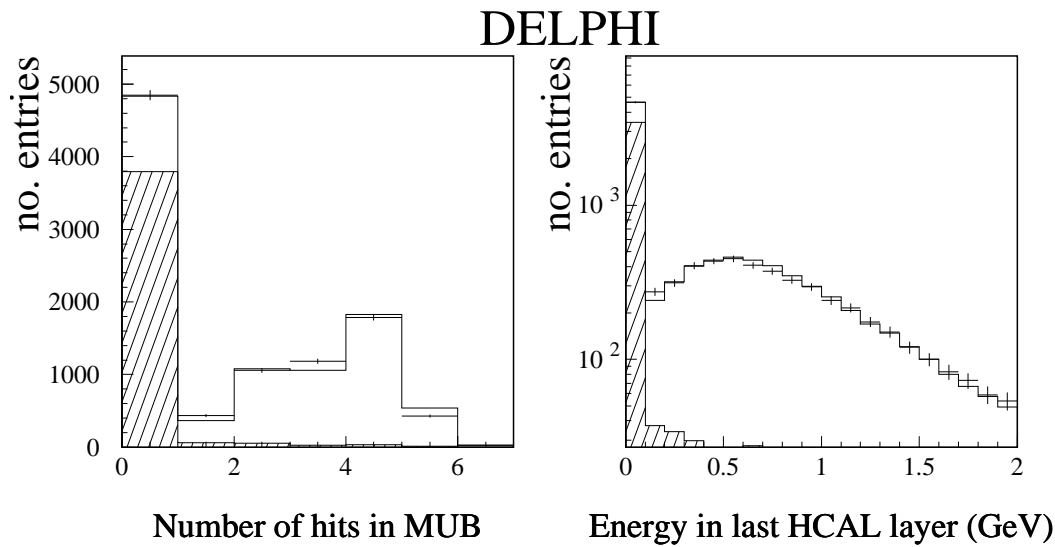}
 \caption{The number of hits in the muon chambers and the energy deposited in
   the last layer of the HCAL after application of all the other
   selection cuts except the one shown for 1993 data. The crosses are
   the data, the solid histogram is the sum in simulation of the signal and
   background and the shaded area is the background from $\tau
   \not\rightarrow \mu \nu_\tau \bar{\nu}_{\mu}$ events.}
 \label{fig:vars_muo}
\end{figure}

Positive muon identification required that the particle
deposited energy deep in the HCAL or had a hit in the muon chambers.
This was achieved specifically in the first instance by insisting that
the average energy per HCAL layer $E_{hlay}$ be less than 2 GeV. 
A logical ``OR'' of two variables was also used in the selection. The
track was required to either have a maximum deposited energy in any
HCAL layer of less than 3 GeV together with deposited energy greater
than 0.2 GeV in the last HCAL layer, or have at least one hit in the
muon chambers. This combination of cuts gave a reasonably constant
efficiency over the whole momentum range. The two selection variables,
the energy deposited in the last HCAL layer and the number of hits in
the muon chambers, can be seen in Fig.~\ref{fig:vars_muo}.
The background was suppressed further by requiring that the sum of the
energies of all the electromagnetic neutral showers in an $18^\circ$
cone around the track did not exceed 2 GeV. This cut was effective in
further suppressing $\tau \rightarrow \pi (n \pi^0)$ and $\mathrm{e^+
  e^-} \rightarrow \mu^+ \mu^- \gamma$ events.

The identification criteria were studied on test samples of real data.
The efficiencies of the HCAL and muon chamber cuts were tested across
the whole momentum range by exploiting the redundancy of the two.
After correcting the simulated data for a discrepancy in the depth 
of the energy deposition by hadrons in the HCAL the data were found
to be well described.

Backgrounds arising from non-$\tau$ sources consisted mainly of
$\mathrm{e^+ e^-} \rightarrow \mu^+ \mu^-$, $\mathrm{e^+ e^-
  \rightarrow e^+ e^-} \mu^+ \mu^-$, $\mathrm{e^+ e^- \rightarrow e^+
  e^-} \tau^+ \tau^-$ and cosmic ray events. The $\mathrm{e^+ e^-}
\rightarrow \mu^+ \mu^-$ background was suppressed by the standard
preselection cut, i.e.~$p_{rad}<1$.  The remaining background was
further suppressed by demanding that the event was rejected if there
was an identified muon in each hemisphere with momentum greater than
$0.8E_{beam}$ and the total visible energy was greater than $70\%$ of
the centre-of-mass energy. The event was also rejected if the momentum
of the identified muon was greater than $0.8p_{beam}$ and the momentum
of the leading track in the opposite hemisphere was greater than
$0.8p_{beam}$.
%To suppress remaining cosmic
%background it was required, if the opposite hemisphere 
%contained an identified muon candidate, that at least 
%one of the  charged particle tracks extrapolated 
%to within 0.3~cm in the $r-\phi$ plane of the interaction region.

The four-fermion events $\mathrm{e^+ e^- \rightarrow e^+ e^-}
\mu^+\mu^-$ and $\mathrm{e^+ e^- \rightarrow e^+ e^-} \tau^+\tau^-$,
although background processes, required no further suppression.

Candidate $\tau
\rightarrow \mu \overline{\nu}_{\mu} \nu_{\tau}$ decays with
muon momenta below 2~GeV/c were selected with different criteria. 
At these energies muons do not have sufficient
energy to penetrate through the HCAL to reach the muon chambers, thus
making the selection more difficult.  Instead, at these lower momenta,
muon candidates were selected if the particle was seen in the last 3
layers of the HCAL.  This procedure was tested using a sample of
hadrons selected from the data and simulation by tagging $\rho$
decays through the presence of a $\pi^0$ in the HPC.  In order to
study the signal, various variables were compared in the data and
simulated data to see if the simulation correctly modelled the
performance of DELPHI at these low energies.  The response of the HCAL
to these hadrons and muons with momenta below 2~GeV/c was well
described by the simulation, after the correction described above
for the hadronic showers.

As a result of the above procedure $\sim 26000$ $\tau \rightarrow \mu
\overline{\nu}_{\mu} \nu_{\tau}$ candidates were selected from the
1992 to 1995 data. The efficiency of selection within the $4\pi$
angular acceptance was $45\%$, the background arising from $\tau
\not\rightarrow \mu \overline{\nu}_{\mu} \nu_{\tau}$ processes was
estimated to be $(1.88\pm0.56)\%$, from $\mathrm{e^+e^-} \rightarrow
\mu^+ \mu^-$ events $(0.52\pm0.16)\%$, from $\mathrm{e^+e^-
  \rightarrow e^+e^-} \mu^+ \mu^-$ events $(0.58\pm0.17)\%$, from
$\mathrm{e^+e^- \rightarrow e^+e^-} \tau^+ \tau^-$ events
$(0.48\pm0.14)\%$ and from cosmic-rays $(0.14\pm0.04)\%$.

\boldmath
\subsection{The $\tau \rightarrow h (n \pi^0) \nu_{\tau}$ channel}
\unboldmath
The $\tau \rightarrow$ inclusive one-prong hadrons channel makes no
distinction between the primary semi-leptonic decays namely $\tau
\rightarrow \pi \nu_{\tau}$, $\tau \rightarrow \rho \nu_{\tau}$ and
$\tau \rightarrow a_1 \nu_{\tau}$.  Instead each decay candidate is
separated into bins of invariant mass, constructed from the 4-momenta
of the charged particles and all reconstructed photons. The invariant
mass bins used were $M_{inv}<0.3$~$\mathrm{GeV/c^2}$,
$0.3$~$\mathrm{GeV/c^2}$~$<M_{inv}<0.95$~$\mathrm{GeV/c^2}$ and
$M_{inv}>0.95$~$\mathrm{GeV/c^2}$.

The preselection of the $\tau$'s for this channel is slightly
different to that for the leptonic channels due to the smaller
potential backgrounds arising from di-lepton events. Therefore there
is no $p_{rad}$ cut in the preselection and the $E_{rad}$ cut is
loosened to~1.1.

In order to identify hadrons one is forced to use almost all the components of
the detector. To be identified as a hadron it was required that one particle 
was detected in a given hemisphere in the angular range 
$0.035 < \left| \cos{\Theta} \right| < 0.732$.
In the case of more than one particle being detected, the hemisphere was 
retained if the highest momentum particle was the only particle having 
associated vertex detector hits. This ensured that one also 
retains a high efficiency for one-prong $\tau$ decays containing 
conversions within the detector.

\begin{figure}[hbt]
\centering
 \epsfxsize=7.0cm
 \leavevmode
 \epsfbox[70 160 560 670]{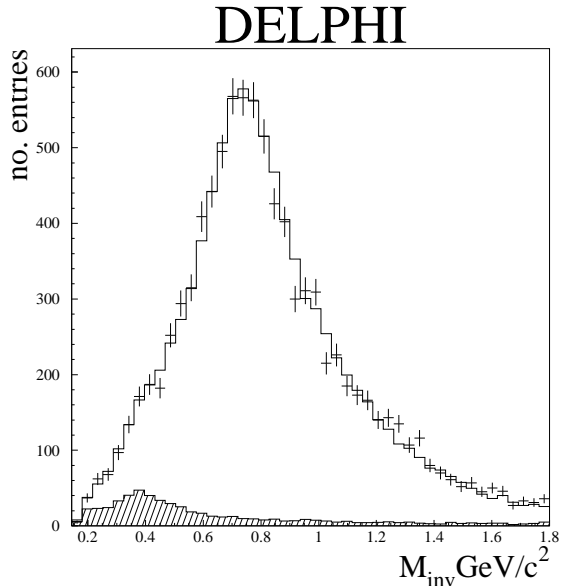}
 \caption{The invariant mass distribution for all
   preselected $\tau$ decays in 1995 data. The crosses are the real
   data, the solid histogram is the simulated $\mathrm{e^+ e^-}
   \rightarrow \tau^+ \tau^-$ data together with the simulated
   background, the shaded area is the sum of the $\mathrm{e^+ e^-
     \rightarrow e^+ e^-}$, $\mathrm{e^+ e^-} \rightarrow \mu^+ \mu^-$
   and the leptonic $\tau$ decays. The pole at the $\pi^{\pm}$ mass is
   not plotted.}
\label{fig:m_inv}
\end{figure}

Further cuts were made depending on the invariant mass of the decay
products.  Fig.~\ref{fig:m_inv} shows the invariant mass distribution
for all preselected $\tau$'s, calculated assuming that all charged
particles were pions and all neutrals were photons. 
Most background from leptons comes at low invariant mass. Hence
one should apply stricter criteria for these events.

The background from electrons was suppressed with the following
two cuts. Firstly the measured $dE/dx$ in the TPC had to be
consistent with being a pion, so ${\prod^{\pi}}_{dE/dx} < 2$.
Because of the importance of the $dE/dx$ measurement to the
selection it was also required that there were at least 28 anode
wires with an ionisation measurement. This cut is particularly
effective at low momentum.

The second cut required that either the particle deposited an energy
beyond the first layer of the HCAL or that the associated energy in
the first four layers of the HPC be less than 1 GeV for invariant
masses below 0.3~$\mathrm{GeV/c^2}$, and 5 GeV otherwise. This cut is
particularly effective at high momentum. The combination of the two
cuts therefore leads to an even efficiency for the suppression of
electrons across the whole momentum range.

Rejection of background from muons was only performed for events with
invariant masses less than 0.3~$\mathrm{GeV/c^2}$. Muon background in
higher invariant mass bins was found to be small enough to justify no
further suppression. The muon rejection was based on the average
energy per HCAL layer, $E_{hlay}$. It was required that either $E_{hlay}$ was
greater than 2 GeV or that there was no energy deposited in the HCAL.
% updated above by PS 13-jan-99
%It was required that this was either zero or greater than 2 GeV. 
In addition to this criterion it was
also required that there were no hits in the muon chambers and that
the momentum of the leading charged particle was greater than
$0.05p_{beam}$ in order that it had sufficient energy to reach the
muon chambers. For regions not covered by the muon chambers it was
required that there was no deposition in the last two layers of the
HCAL. In this instance any tracks pointing to HCAL azimuthal
boundaries were rejected.

The identification criteria were studied with test samples of real
data. The efficiencies of all the main selection cuts were tested
using a sample of hadrons selected by tagging $\pi^0$'s in the HPC.
This test sample allowed for an accurate calibration of all the main
selection variables across the whole range of $\cos\theta^*$ and $\cos
\psi$, the two variables used in the fits to the Michel parameters and
the anomalous tensor coupling.

Remaining background from $\mathrm{e^+ e^- \rightarrow e^+ e^-}$ and
$\mathrm{e^+ e^-} \rightarrow \mu^+ \mu^-$ events was suppressed by
demanding that the particle in the opposite hemisphere to the
identified hadron had a measured momentum of less than $0.8p_{beam}$.
The four-fermion events $\mathrm{e^+ e^- \rightarrow e^+ e^-}
\tau^+\tau^-$ required no further suppression.

A total of $\sim 56000$ $\tau \rightarrow h (n
\pi^0) \nu_{\tau}$ candidates were selected from the data. The
efficiency of selection within the $4\pi$ angular acceptance was
$37\%$, the background arising from $\tau \not\rightarrow h (n \pi^0)
\nu_{\tau}$ processes was estimated to be $(2.43\pm0.73)\%$ from
$\mathrm{e^+ e^- \rightarrow e^+ e^-}$ events, $(0.40\pm0.12)\%$ from
$\mathrm{e^+ e^-} \rightarrow \mu^+ \mu^-$ events $(0.10\pm0.03)\%$
and from $\mathrm{e^+ e^- \rightarrow e^+ e^-} \tau^+ \tau^-$ events
$(0.23\pm0.07)\%$.

\subsection{The two-dimensional selection}

As described in Section~\ref{mich}, in order to measure the Michel
parameters most efficiently it is necessary to use two-dimensional
spectra.  It was required that the events satisfied the preselection
cuts and that there was one identified candidate $\tau$ decay in each
hemisphere. This therefore produces 20 (15 two-dimensional and 5
one-dimensional) distributions consisting of e$\mu$, ee, $\mu\mu$,
e$h1$,\footnote{where h1,h2 and h3 are hadrons in the invariant mass
  bins $M_{inv} < 0.3$ $\mathrm{GeV/c^2}$, $0.3$ $\mathrm{GeV/c^2} <
  M_{inv} < 0.95$ $\mathrm{GeV/c^2}$ and $M_{inv} > 0.95$
  $\mathrm{GeV/c^2}$ respectively} e$h2$, e$h3$, $\mu h1$, $\mu h2$,
$\mu h3$, $h1h1$, $h2h2$, $h3h3$, $h1h2$, $h1h3$, $h2h3$, e$X$, $\mu
X$, $h1X$, $h2X$ and $h3X$, where the two identified particles in each
correspond to the two hemispheres in the event. The $X$ in the event
is an unidentified $\tau$ decay with either one or three charged
particles. In this case only the hemisphere with the identified track
is used.

In most of these channels it is required that the $\tau$
preselection cuts be satisfied in order that non-$\tau$
backgrounds be suppressed. This is not true for the $e\mu$ channel
in which no preselection cuts were necessary as the external
background required no further suppression. To suppress remaining cosmic ray
background in the $\mu\mu$ and
the $\mu X$ samples it was required, in one-versus-one charged particle 
topologies, that at least 
one of the  charged particle tracks extrapolated to within 0.3~cm
in the $r-\phi$ plane of the interaction region.
For the one-dimensional distributions, e$X$, $\mu X$,
$h1X$, $h2X$ and $h3X$, the cuts to remove external backgrounds
follow those already outlined in the previous sections describing
the one-dimensional selections.

The number of events selected, the efficiency of selection within the fiducial
volume and momentum acceptance and the backgrounds can be seen in 
Tables~\ref{t:eff2d} and~\ref{t:bac2d}.
\begin{table}[hbtp]
\begin{center}
 \begin{tabular}{l|c}
  \hline
  \multicolumn{1}{c|}{channel} & efficiency(\%) \\
  \hline
  $Z^0 \rightarrow \tau^+ \tau^- \rightarrow
  (\mathrm{e} \nu \overline{\nu})(\mu \nu \overline{\nu})$
     & $72.95\pm0.23$  \\
  $Z^0 \rightarrow \tau^+ \tau^- \rightarrow
  (\mathrm{e} \nu \overline{\nu})(\mathrm{e} \nu \overline{\nu})$
     & $50.43\pm0.36$  \\
  $Z^0 \rightarrow \tau^+ \tau^- \rightarrow
  (\mu \nu \overline{\nu})(\mu \nu \overline{\nu})$
     & $82.77\pm0.27$  \\
  $Z^0 \rightarrow \tau^+ \tau^- \rightarrow
  (\mathrm{e} \nu \overline{\nu})(h (n \pi^0) \nu)$
     & $47.08\pm0.15$  \\
  $Z^0 \rightarrow \tau^+ \tau^- \rightarrow
  (\mu \nu \overline{\nu})(h (n \pi^0) \nu)$
     & $60.23\pm0.15$  \\
  $Z^0 \rightarrow \tau^+ \tau^- \rightarrow (h (n \pi^0) \nu)(h (n \pi^0) \nu)$
     & $37.62\pm0.13$  \\
  \hline
 \end{tabular}
\caption{The efficiencies of selection in the angular and momentum
acceptance for the two-dimensional analysis in the 1994 data set.
The efficiencies were similar for the other years.
The errors are purely statistical.} \label{t:eff2d}
\end{center}
\end{table}
\begin{table}[hbtp]
\begin{center}
 \begin{tabular}{c|c|c|c}
  \hline
   \multicolumn{1}{c|}{$\tau^+\tau^-$ decay}
 & \multicolumn{1}{c|}{no. of candidate} 
 & \multicolumn{1}{c|}{internal}
 & \multicolumn{1}{c}{non-$\tau^+\tau^-$}  \\
   \multicolumn{1}{c|}{modes}
 & \multicolumn{1}{c|}{events} 
 & \multicolumn{1}{c|}{background(\%)}
 & \multicolumn{1}{c}{background(\%)}  \\
  \hline
    e-e & 1405 & $6.01\pm1.80$ & $7.19\pm2.16$ \\
    e-$\mu$ & 3495 & $4.43\pm1.33$ & $0.60\pm0.18$ \\
    e-$h1$& 1804 & $5.31\pm1.59$ & $0.60\pm0.18$ \\
    e-$h2$ & 3324 & $3.92\pm1.18$ & $0.11\pm0.03$ \\
    e-$h3$ & 1088 & $4.01\pm1.20$ & $0.21\pm0.06$ \\
    e-$X$ & 6377 & $2.96\pm0.89$ & $4.66\pm1.40$ \\
    $\mu$-$ \mu$ & 2116 & $2.61\pm0.78$ & $3.89\pm1.17$ \\
    $\mu$-$h1$   & 2160 & $3.88\pm1.16$ & $0.30\pm0.09$ \\
    $\mu$-$ h2$ & 4454 & $2.13\pm0.64$ & $0.54\pm0.16$ \\
    $\mu$-$h3$ & 1480 & $2.09\pm0.63$ & $0.80\pm0.24$ \\
    $\mu$-$X$ & 8632 & $1.58\pm0.47$ & $1.36\pm0.41$ \\
    $h1$-$h1$ & 571 & $4.89\pm1.47$ & $0.20\pm0.06$ \\
    $h1$-$h2$    & 2271 & $3.19\pm0.96$ & $0.10\pm0.03$ \\
    $h1$-$h3$ & 730 & $3.12\pm0.94$ & $0.15\pm0.05$ \\
    $h1$-$X$ & 5104 & $2.54\pm0.76$ & $1.87\pm0.56$ \\
    $h2$-$h2$ & 2295 & $1.69\pm0.51$ & $0.12\pm0.04$ \\
    $h2$-$h3$ & 784 & $1.80\pm0.54$ & $0.01\pm0.01$ \\
    $h2$-$X$ & 9342 & $0.93\pm0.28$ & $0.33\pm0.10$ \\
    $h3$-$h3$ & 278 & $1.93\pm0.58$ & $0.01\pm0.01$ \\
    $h3$-$X$ & 3058 & $0.94\pm0.28$ & $0.16\pm0.05$ \\
  \hline
%    Total & 60768 &  \\
%  \hline
 \end{tabular}
\caption{The number of selected events (column 2) and backgrounds
  (columns 3 and 4) for the selection described in 
  the text. The backgrounds are quoted for the 1994 data set only.
  They were similar for the other years. A total of
  60768 events were selected in the 1992-1995 sample.}
\label{t:bac2d}
\end{center}
\end{table}

\section{The extraction of the Michel parameters}
\label{sec:results}

The  values  of  the  Michel  parameters, $\rho$,  $\eta$,  $\xi$  and
$\xi\delta$ together with the tau polarisation, ${\cal P}_{\tau}$, and
the tau  neutrino helicity,  $h_{\nu_{\tau}}$, are extracted  from the
data using a binned maximum  likelihood fit to all the combinations of
$\tau  \rightarrow   \mathrm{e}  \overline{\nu}_e  \nu_{\tau}$,  $\tau
\rightarrow \mu \overline{\nu}_{\mu} \nu_{\tau}$ and $\tau \rightarrow
h (n  \pi^0) \nu_{\tau}$.  In splitting the  hadron sample  into three
invariant  mass bins  one is  left  with 15  two-dimensional and  five
one-dimensional  distributions where  only one  $\tau$ decay  has been
exclusively identified in an event.

The likelihood function is defined as:
\begin{equation}
\mathcal{L} = \mathit{\prod_{c} \prod_{i,j}
                        \frac { ({a^c_{ij}})^{(n^c_{ij})} e^{-a^c_{ij}} }
                                        { (n^c_{ij}) ! } }
\end{equation}
where $n^c_{ij}$ is the number of observed events
in selected class $c$ in the bin denoted by the indices $i,j$. 
The predicted number of events in this bin is $a^c_{ij}$ 
and is given by
\begin{equation}
a^c_{ij} = {\cal E}^c_{ij} \; \sum_{i',j'} 
  \; {\cal T}^c_{i'j'} \; {\cal R}^c_{i'i} \; {\cal R}^c_{j'j} \;
  +\; b^{c,\tau}_{ij} \;
  +\; b^{c,\mathrm{non-}\tau\tau}_{ij}.
\end{equation}
The detector resolution matrix
${\cal   R}^c_{k'k}$  gives  the  fraction  of
reconstructed signal  events with generated  fit variable in  bin $k'$
which are  reconstructed in bin  $k$. ${\cal E}^c_{ij}$  describes the
$\tau^+\tau^-$ selection efficiency as a function of the reconstructed
fit variables in the two  $\tau$ decay hemispheres. The ${\cal  R}$ 
and  ${\cal E}$ matrices  were obtained from  the full
detector  simulation.  The matrix ${\cal
T}$   contains  the two-dimensional  distribution corresponding 
to Eqn.~\ref{eqn:2d_dist} and   the
dependence on  the fitted parameters. The construction  of ${\cal T}$,
taking into account mass, radiation, and hadronic modelling effects, is
described below.  The number of background $\tau\tau$ events per bin is
$b^{c,\tau}_{ij}$,  and was  not varied  as a  function of  the fitted
parameters.      The     non-$\tau\tau$     background    per     bin,
$b^{c,\mathrm{non-}\tau\tau}_{ij}$,  was normalised to  the luminosity
of the  data. The  signal and $\tau$  background were  then normalised
keeping  their ratio constant  so that  the integrals  of the
predicted  fit distributions  were the  same  as the  total number  of
events seen in the data.

This  method  accounted  for  correlations between  the  $\tau^+$  and
$\tau^-$ in  an event  arising from geometric  detector reconstruction
effects, described by the detector simulation, and  physical
effects such  as longitudinal  spin correlations, electroweak  and QED
corrections,  described by  the KORALZ  program.  Near  the  $Z$ pole,
photonic radiative effects are a strong function of the centre-of-mass
energy.  In the derivation  of the efficiency and resolution matrices,
$\tau^+\tau^-$  simulation samples  have been  used for  the different
centre-of-mass  energies with  proportions corresponding  to  the data
sample.

The $h_i(x)$ polynomials describing  the leptonic decay spectrum shown
in  Fig.~\ref{fig:polys} do  not  take into  account  mass effects  or
radiative corrections.   These effects were introduced  by Monte Carlo
methods  using KORALZ  and  a modified  version~\cite{tauola2} of  the
TAUOLA program to generate distributions corresponding to the $h_i(x)$
polynomials.  The  TAUOLA program  models leptonic $\tau$  decays with
the   matrix   element  containing   exact   ${\cal  O}(\alpha)$   QED
corrections.  The  modified version contained a  generalisation of the
Born level part  of the matrix element which  permitted the setting of
non-Standard Model values for the  Michel parameters.  The part of the
matrix element describing the  QED corrections was calculated assuming
$V-A$  couplings.  The part  of  the  matrix  element proportional  to
$\alpha_{QED}$  is  small  and   it  was  assumed  that  for  observed
variations of the  Michel parameters the change in  the spectra due to
changes in the radiative corrections could be neglected.

For semi-leptonic $\tau$ decays,  the distributions were obtained from
linear     combinations     of     distributions    generated     with
$h_{\nu_{\tau}}=-1$ and either positive or negative helicity states of
the decaying~$\tau$.

In the fit it was assumed that $h_{\nu_{\tau}}$ had the same value for
all the semi-leptonic decay modes.

In the  fit assuming  lepton universality the  value of $\eta$  can be
constrained using the measured values of the leptonic branching ratios
in  Eqn.~\ref{eqn:etabr}.  The  branching  ratio results~\cite{bjarne}
were obtained  from the  DELPHI data in  the years 1991  through 1995.
The value of $\eta$ was constrained with the addition of the following
quantity to the log-likelihood function
\begin{eqnarray}
\ln {\cal L^{\rm const}} & = & - \frac 1 2
\frac { ( \eta - \eta_{Br} )^2 } { (\Delta \eta_{Br})^2 },
\end{eqnarray}
where $\eta_{Br}$ is the value
obtained from the leptonic branching ratio measurement and $\Delta
\eta_{Br}$ is the error on this measurement.

It must however be noted that obtaining 
Eqn.~\ref{eqn:etabr} involves an integration over the final
state momenta, the implications of which have to be accounted for
when setting a limit on $\eta$ based on experimentally measured
branching fractions. Since $\eta$ affects the shape of the muon
momentum spectrum as well as the total decay rate, it is necessary
to study the effect of the cutoff on the muon momentum
identification which is at $ x^c = p^c/ p_{beam} = 0.05$. As a
function of the normalised laboratory muon momentum $x = p / p_{beam}$ 
the number of events observed between momentum $x$ and
$x + dx$ can be written as
\begin{equation}
\label{eqn:xdis}
  d N = N_0 \left[ a(x) + K \eta  b(x) \right]dx .
\end{equation}
By analogy with Eqn.~\ref{eqn:leptdecay2}, the polynomial 
$a(x) \equiv h_0(x) + \textstyle\frac{3}{4}h_\rho(x) - {\cal P}_{\tau}
[h_\xi(x)+\textstyle\frac{3}{4}h_{\xi\delta}(x)]$ 
is the appropriate linear combination
of polynomials for Standard Model couplings at LEP energies, while
$b(x) \equiv h_\eta(x)$.
%The polynomials $a(x), b(x)$ are functions of the readily
%obtainable decay distribution expressed in the $\tau$ rest frame
%as a function of energy, decay angle, $\tau$ polarisation and the
%other Michel parameters, with well defined values in the Standard
%Model~\cite{stahl,pich}. 
The constants $N_0$ and $K$
can always be chosen such that the integrals of $a(x)$ and $b(x)$
over the whole momentum range are normalised to 1. If $\eta$ is
non-zero, the number of events observed would be
\begin{equation}
\label{eqn:nobs} N_{obs} = N_{0} \left[ \int_{x^c}^{x^{max}} a(x)
dx +
                K \eta  \int_{x^c}^{x^{max}} b(x) dx
             \right] .
\end{equation}
\par
The event generator used to compute the acceptance corrections, 
KORALZ/TAUOLA, assumes that $\eta$ equals zero. In other words, the 
branching ratio is derived assuming that the total number of 
$\tau\to\mu \overline{\nu}_{\mu} \nu_{\tau}$ 
decays produced can be estimated as
\begin{equation}
\label{eqn:acor} N_{0}^{est} = N_{obs} \times \frac{ 1}
                            {\int_{x^c}^{x^{max}} a(x)  dx}
\end{equation}
where the integral is obtained from simulation. Hence, instead of
correcting to obtain $N_{0}^{est} = N_{0} + K \eta$, the estimate of
the corrected number of events becomes
\begin{equation}
\label{eqn:n0} N_{0}^{est}  = N_{0} \left[ 1 +
            K \eta \frac{ \int_{x^c}^{x^{max}} b(x) dx}
                 {\int_{x^c}^{x^{max}} a(x) dx}
           \right] .
\end{equation}
\par
The ratio between the integrals is readily calculated numerically
by generating the full distribution in the $\tau $ rest frame and
boosting  the momentum to the lab frame. It is found that  the
ratio between the integrals equals 0.96 when integrating from $x^c
= 0.05$. Ignoring effects due to $\eta$ in 
$\tau\to \mathrm{e \overline{\nu}_{e} \nu_{\tau}}$ 
decays, the relation
\begin{eqnarray}
\frac {Br ( \tau \rightarrow \mu \nu_{\tau} \overline{\nu}_{\mu}
)}
      {Br ( \tau \rightarrow \mathrm{e} \nu_{\tau} \overline{\nu}_e )}
       & = &
f \biggl( \frac {m_{\mu}^2} {m_{\tau}^2} \biggr) + 3.84 \frac
{m_{\mu}} {m_{\tau}} g \biggl( \frac{m_{\mu}^2}{m_{\tau}^2}\biggr)
\eta_{\mu} . \label{eqn:eta}
\end{eqnarray}
should be used to extract $\eta_{\mu}$ from the DELPHI  tau leptonic
branching ratios instead of Eqn.~\ref{eqn:etabr}.
%
%The result
%$$ \eta =  -0.025 \pm 0.039 \pm 0.031 $$ is obtained for the full
%1991-1995 dataset.
%
%   The figure (Momentum distributions for eta=0 and eta=1
%
%\begin{figure}[tbph]
%\vspace{0.1cm}
%\begin{center}
%\mbox{\epsfxsize 15.0cm\epsfbox{etapl.ps}}
%\end{center}
%\vspace{-2.0cm} \caption {   Generator level momentum
%distributions in the lab frame for two values of the parameter
%$\eta$ in $\tau\to\mu$ decays. No radiative corrections are applied}
%\label{fig:etapl}
%\end{figure}
%

\begin{figure}[hbt]
\centering   \epsfxsize=14.0cm   \leavevmode   \epsfbox[19   150   526
  669]{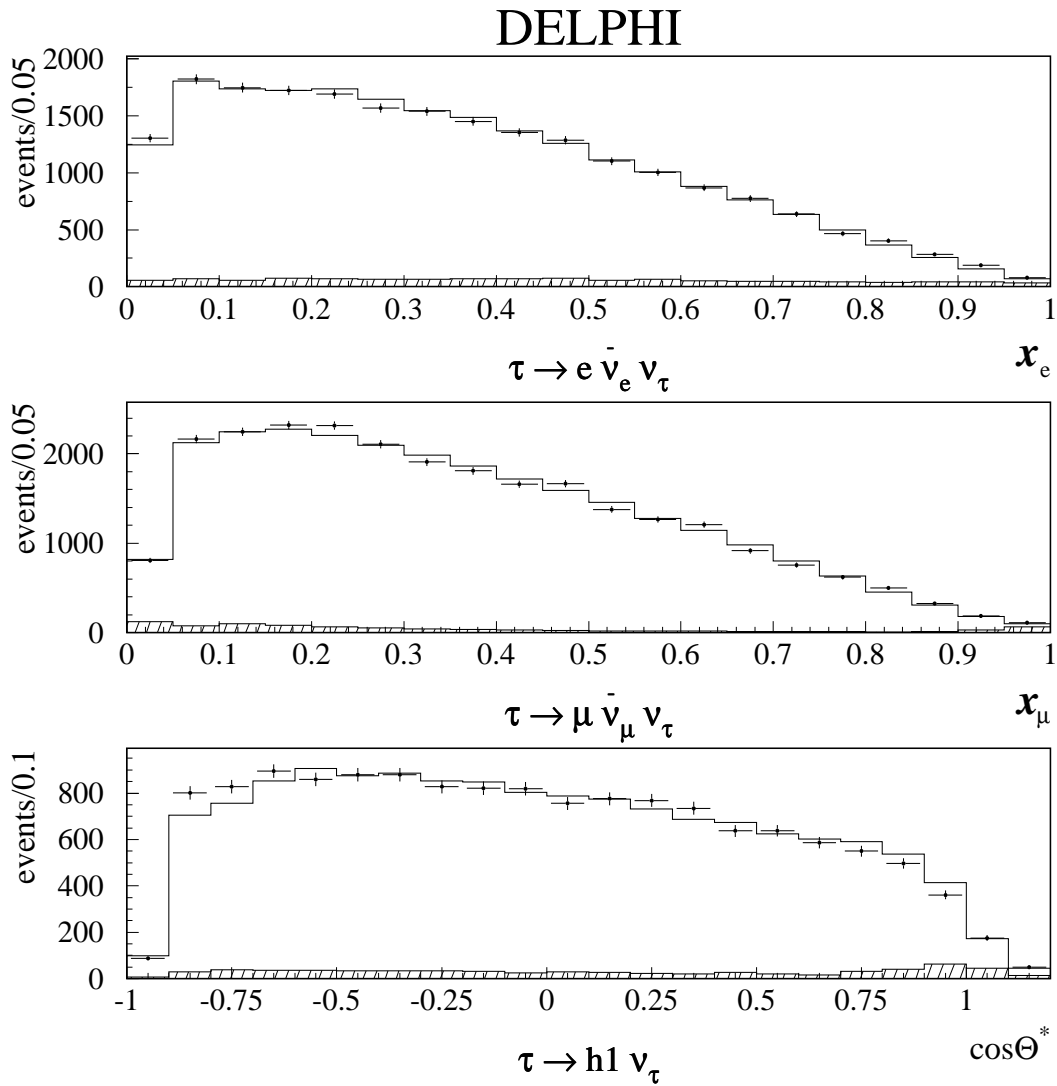}
 \caption{The projections of the 
   fitted distributions for the six parameter fit for the two fully
   leptonic decay  channels and the semi-leptonic  candidates from the
   lowest invariant mass  bin. The line is the result  of the fit, the
   points  are  the  data and  the  shaded  area  is  the sum  of  the
   backgrounds.}
 \label{fig:bfit1}
\end{figure}

\begin{figure}[hbt]
\centering   \epsfxsize=14.0cm   \leavevmode   \epsfbox[17   386   535
  679]{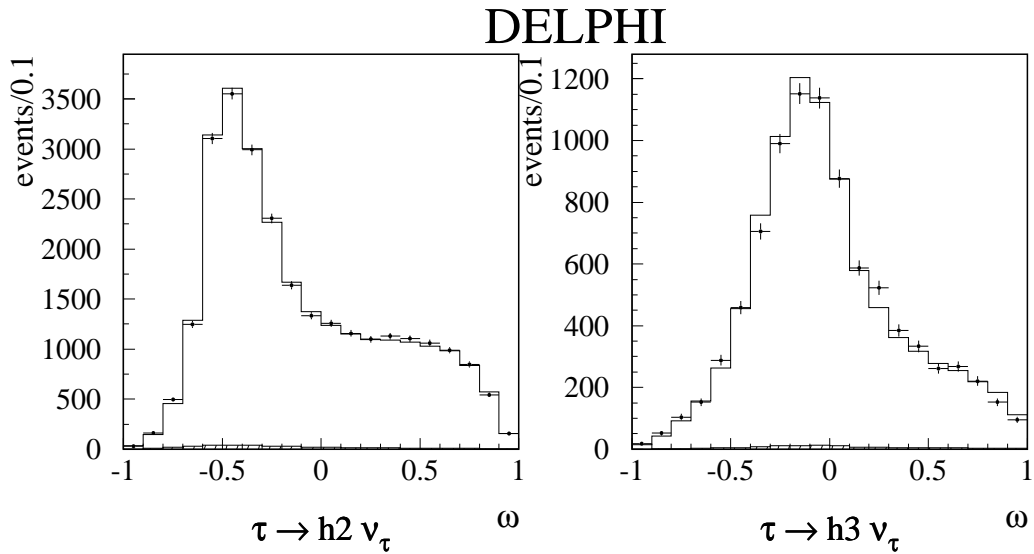}
 \caption{The projections of the
 fitted distributions of the $\omega$ variable described 
 in~\cite{rouge}  from the  six  parameter fit  for the  semi-leptonic
 candidates selected in the second  and third invariant mass bins. The
 line is the result of the fit, the points are the data and the shaded
 area is the sum of the backgrounds.}
 \label{fig:bfit2}
\end{figure}

Using   the  techniques  outlined   above  together   with  background
distributions obtained from  the simulated data a six  parameter and a
nine parameter fit were performed,  with and without the assumption of
lepton universality respectively, over  a sample of $\sim$60000 $\tau$
pair candidates.  The one-dimensional projected distributions for each
$\tau$ decay class are shown in
Figs.~\ref{fig:bfit1}~and~\ref{fig:bfit2}, together with the 
fitted distributions obtained from the six parameter fit.        

%The       systematic
%uncertainties  on the  measurement  arose from  the  finite amount  of
%simulated  data  available, the  uncertainties  in  the world  average
%values of the $\tau$ branching ratios, the uncertainties in the levels
%of the  backgrounds, uncertainties in the efficiency  of selection and
%calibration uncertainties on the  selection and fit variables. For the
%particular case  of the fit  assuming lepton universality there  is an
%additional systematic  arising from the uncertainties  on the leptonic
%branching  ratio  measurement.   The  systematics  are  summarised  in
%Tables~\ref{t:sysuni}    and~\ref{t:sysnouni}.    

A number of cross-checks were  performed to check the stability of the
result with respect to the selection cuts and binning effects. Each of
the $\tau^+\tau^-$ preselection  cuts was varied by 10\%  of its value
(in the case of the  $E_{rad}$ and $p_{rad}$ by 5\% corresponding more
closely to their resolution); no variation in the results was observed
beyond those expected from statistical fluctuations. A similar process
was  performed for  the  main cut  criteria  in the  selection of  the
different $\tau$ decay modes  classes. Again no variation was observed
in the fit results beyond that expected from statistical fluctuations.
The binning used  to define the hadronic decay  classes $h1$, $h2$ and
$h3$ was varied; no unexpected variation in the results was observed.

The systematic  effects studied are  decribed below and  summarised in
Tables~\ref{t:sysuni} and~\ref{t:sysnouni}.

One source of systematic  uncertainty  arose from  the finite
amount of simulated data available.

An  uncertainty due  to the  $\tau$ branching  ratios was  obtained by
varying the branching ratios by the uncertainties on the world average
values, repeating  the fit and taking  the change in the  result as an
estimate   of  the   systematic   uncertainty.   Conservatively,   the
background from $\tau$ decays  and the $\tau^+\tau^-$ backgrounds were
varied by 30\%  and the change in  the results of the fit  taken as an
estimate of the systematic uncertainty.

The dependence  of the selection  efficiency on the fit  variables for
the different $\tau$  decay modes was studied using  data test samples
or by redundancy of the  different detector components as described in
each of the relevant sections.  The resulting systematic uncertainties
were estimated  by varying the selection efficiency  in the simulation
as  a linear  function  of the  fit  variable.  The  magnitude of  the
variation was  taken from the statistical uncertainty  on the gradient
derived  in  a  straight  line  fit  to  the  ratio  of  the  measured
efficiencies in data and simulation as a function of the fit variable.

Systematic  uncertainties  were  attributed for  detector  calibration
effects.   The  charged  particle  reconstruction momentum  scale  was
varied by  its uncertainty, the  analysis repeated, and  the resultant
variation in the results taken  as an uncertainty.  This in particular
affected the muon channel parameters.  The effects due to knowledge of
the momentum  resolution were  also taken into  account but  were much
smaller.

The electron momentum estimator  $p_{el}$ was calibrated on data using
both  radiative  and  non-radiative  Bhabha  events.   Its  scale  was
calibrated with a precision of 0.5\%, limited by the statistics of the
data test  samples.  The systematic  uncertainties on the  various fit
parameters were estimated  in the same way as  for the momentum scale.
A smaller contribution  arose from the knowledge of  the resolution on
$p_{el}$.

The neutral electromagnetic energy scale was known with a precision of
0.2\%.   The  related uncertainties  were  estimated  in an  analogous
manner  to  those due  to  the  momentum  and $p_{el}$  scales.   This
affected mostly the  hadronic decay modes with $\pi^0$'s  in the final
state.

The  uncertainty in  the energy  scale in  the HCAL  had  a negligible
effect, as did the uncertainty in the efficiency of the muon chambers.

The systematic uncertainty contribution arising from the HCAL response
to hadronic showers was  estimated by varying within their statistical
errors  the corrections to  the shower  penetration in  the simulation
taken from the data test samples.

The  hadronic invariant  mass scale  uncertainty is  dominated  by the
neutral  energy and charged  particle momentum  scale.  
Any serious discrepancy between simulation and data
would be evident  in the hadronic invariant mass  distribution such as
that   shown  in   Fig~\ref{fig:m_inv},  where   agreement   is  good.
Additional  checks have  been made  on the  spatial resolution  of the
electromagnetic showers  in the HPC.   These effects were found  to be
small compared with those due to energy scale and resolution.

Imperfections in the modelling of the photon reconstruction efficiency
could  lead to  a  poorly modelled  cross-talk  between the  different
invariant mass classes in the  inclusive hadronic selection as well as
affecting the reconstruction of  the $\theta^*$ and $\psi$ angles used
in  the  fit.    From  a  study~\cite{delphipolarisation}  of  various
distributions related  to reconstructed photons,  such as multiplicity
and  energy distributions, it  was estimated  that the  neutral photon
reconstruction efficiency was known  to better than 4\% averaging over
the  whole of the  HPC taking  into account  dead space  and threshold
effects.  The systematic uncertainty  attributed to this was estimated
by randomly rejecting 4\% of photons and the change in the results was
included under  the heading  calibration. Further cross-checks  of the
HPC   reconstruction  in  the   inclusive  hadronic   sample  included
reclassifying  energy depositions associated  to the  charged particle
track  as neutral  particles.  This  had a  negligible  effect on  the
results,  indicating  that  both  the mis-association  of  photon  and
$\pi^0$  showers to  the charged  hadron  and the  description of  the
hadronic  interactions  associated to  the  hadron  charged track  and
misidentified as  electromagnetic showers  were well described  by the
simulation.

The  uncertainties due  to  radiative corrections  in hadronic  $\tau$
decays  and modelling  of  the $a_1$  have  been estimated  to give  a
systematic uncertainty of 0.001 on ${\cal P}_{\tau}$ for the inclusive
hadronic polarisation analysis in~\cite{delphipolarisation}.  This has
been included as a systematic  on ${\cal P}_{\tau}$ and the systematic
uncertainty has been propagated through to the other fit parameters.

Fig.~\ref{fig:bfit1}
shows discrepancies between data  and the fitted distributions for the
$\tau \rightarrow  h1 \nu_{\tau}$  sample.  Studies of  the quantities
used  to select  the sample  exhibited no  obvious effect  which could
account for this. A cross-check, performing the fit excluding the data
from the discrepant regions  (the ranges [-0.9,-0.7] and [0.8,1.0]) of
the  $\tau \rightarrow  h1 \nu_{\tau}$,  showed variations  which were
consistent with  statistical fluctuations.  Conservatively, systematic
uncertainties were  estimated for  this effect by  taking half  of the
variation in the  fit results when forcing the fits  to go through the
data  points  in   the  quoted  ranges.   These  are   included  as  a
contribution     to     the     ``calibration''     uncertainty     in
Tables~\ref{t:sysuni} and~\ref{t:sysnouni}.

\begin{table}[hbtp]
\begin{center}
 \begin{tabular}{c|c|c|c|c|c|c}
   \hline 
& $\eta$ & $\rho$ & ${\cal P}_{\tau}$ & $\xi$ & $\xi \delta$ & $h_{\nu_{\tau}}$ \\
  \hline 
MC  stats       & 0.0053  & 0.0035  & 0.0018  &  0.0104 &  0.0103  &  0.0039 \\
$\tau$ BR's     &  0.0002 & 0.0006  &  0.0014 &  0.0004 &  0.0012  &  0.0020 \\
Backgrounds     &  0.0251 &  0.0115 & 0.0011  &  0.0030 &  0.0126  &  0.0093 \\
Efficiency      & 0.0005  & 0.0023  & 0.0037  &  0.0013 &  0.0027  &  0.0014 \\
Calibration     &  0.0144 & 0.0146  & 0.0065  &  0.0281 &  0.0229  &  0.0034 \\
Decay modelling &  -      & 0.0009  & 0.0010  &  0.0007 &  0.0004  &  0.0010 \\
$\eta$ const. &  0.0232 & 0.0070  & -       & -       & -        & -       \\
 \hline 
Total Syst.  & 0.037 &  0.020 & 0.008 &  0.030 &  0.028 & 0.011  \\
 \hline 
Statistical & 0.036 &  0.023 & 0.012 & 0.070 & 0.070 & 0.027 \\
 \hline
 \end{tabular}
\caption{The systematics on the
parameters for the six parameter fit with the assumption of universality. The
statistical error is shown for comparison.}
\label{t:sysuni}
\end{center}
\end{table}
\begin{table}[hbtp]
\begin{center}
 \begin{tabular}{c|c|c|c|c|c|c|c|c|c}
   \hline
       & $\eta_{\mu}$ & $\rho_e$ & $\rho_{\mu}$ & ${\cal P}_{\tau}$ &
       $\xi_e$ & $\xi_{\mu}$ & $\xi_e \delta_e$ & $\xi_{\mu} \delta_{\mu}$ &
       $h_{\nu_{\tau}}$ \\
   \hline
 MC stats &    0.047 & 0.0054 & 0.0144 & 0.0018 & 0.0177 & 0.028 &
 0.019 & 0.019 & 0.0039 \\
 $\tau$ BR's & 0.003 & 0.0008 & 0.0016 & 0.0016 & 0.0017 & 0.001 &
 0.001 & 0.001 & 0.0003 \\
 Backgrounds & 0.138 & 0.0230 & 0.0414 & 0.0013 & 0.0058 & 0.044 &
 0.018 & 0.018 & 0.0093 \\
 Efficiency &  0.010 & 0.0034 & 0.0017 & 0.0036 & 0.0020 & 0.003 &
 0.001 & 0.001 & 0.0014 \\
 Calibration & 0.039 & 0.0278 & 0.0076 & 0.0069 & 0.0438 & 0.018 &
 0.034 & 0.033 & 0.0045 \\
 Decay modelling & 0.002 & 0.0010 & 0.0004 & 0.0010 & -      & 0.001 &
 0.001 & 0.001 & 0.0010 \\
 \hline
 Total Syst. & 0.15 & 0.037 & 0.045 & 0.008 & 0.05 & 0.06 &
 0.04 & 0.04 & 0.011 \\
 \hline
 Statistical & 0.32 & 0.036 & 0.098 & 0.012 & 0.12 & 0.19 &
 0.12 & 0.13 & 0.028 \\
 \hline
 \end{tabular}
\caption{The systematics on the
parameters for the nine parameter fit without the assumption of universality. 
The statistical error is shown for comparison.}
\label{t:sysnouni}
\end{center}
\end{table}

The six parameter fit assuming lepton universality (including the constraint
on $\eta$ from the leptonic branching ratios) gave the following results:
\begin{eqnarray}
\eta & = & -0.005 \pm 0.036 \pm 0.037, \nonumber \\
\rho & = & 0.775 \pm 0.023 \pm 0.020, \nonumber \\
\xi & = & 0.929 \pm 0.070 \pm 0.030, \nonumber \\
\xi \delta & = & 0.779 \pm 0.070 \pm 0.028, \nonumber \\
h_{\nu_{\tau}} & = & -0.997 \pm 0.027 \pm 0.011, \nonumber \\
{\cal P}_{\tau} & = & -0.130 \pm 0.012 \pm 0.008. \nonumber
\end{eqnarray}

The results were found to be stable as a function of the year of data 
taking. After correcting for effects including the
photon propagator and the $\sqrt{s}$ dependence,
the  result on ${\cal P}_{\tau}$ obtained in this analysis can be compared
to the polarisation parameter ${\cal A}_{\tau}$ obtained in the 
dedicated analysis of the DELPHI data~\cite{delphipolarisation}. 
Correcting ${\cal P}_{\tau}$ as described in~\cite{delphipolarisation}
the result ${\cal A}_{\tau} = 0.134 \pm 0.014$ is obtained for this analysis.
This is in excellent agreement with the result 
${\cal A}_{\tau} = 0.1359 \pm 0.0096$ from~\cite{delphipolarisation}.
 
The $\chi^2 / N_{dof}$ for the fit was 1984/2009.
The parameters are correlated and the correlation matrix is given in
Table~\ref{t:corr6}.
\begin{table}[hbtp]
\begin{center}
 \begin{tabular}{c | r r r r r}
   & \multicolumn{1}{c}{$\rho$} 
   & \multicolumn{1}{c}{${\cal P}_{\tau}$} 
   & \multicolumn{1}{c}{$\xi$}
   & \multicolumn{1}{c}{$\xi \delta$}
   & \multicolumn{1}{c}{$h_{\nu_{\tau}}$} \\
 \hline
 $\eta         $ & $0.276$ & $-0.016$ & $0.100$ & $0.070$ & $0.009$ \\
 $\rho         $ &  & $0.435$ &$ -0.060$ & $-0.105$ & $-0.205$ \\
 ${\cal P}_{\tau}$ &  &  & $0.04$0 & $-0.188$ & $-0.414$ \\
 $\xi$           &  &  &   & $-0.142$ & $0.062$ \\
 $\xi \delta  $  &  &  &  &   & $0.190$ \\
 \end{tabular}
 \caption{The correlation matrix for the six parameter fit.}
 \label{t:corr6}
 \end{center}
\end{table}

The variable $P^{\tau}_{R}$, defined in
Eqn.~\ref{eqn:ptaur}, represents the probability of a
right-handed $\tau$ decaying into a lepton of either handedness.
This was calculated~to~be
\begin{eqnarray}
P^{\tau}_{R} & = & -0.038 \pm 0.066 \pm 0.029. \nonumber
\end{eqnarray}
%\begin{figure}[hbt]
%\centering
%  \epsfxsize=14.0cm
%  \leavevmode
%  \epsfbox[17 386 535 679]{bfit2.ps}
% \caption{The fitted distributions of the $\omega$ variable described 
% in~\cite{rouge} from the six parameter fit for the semi-leptonic candidates 
% selected in the second and third invariant mass bins. The line is the result 
% of the fit, the points are the data and the shaded area is the sum of the 
% backgrounds.}  
% \label{fig:bfit2}
%\end{figure}
A one-dimensional fit to $\eta$ was also performed. In setting
the other Michel parameters to their Standard Model values and
applying the branching ratio constraint (Eqn.~\ref{eqn:eta})
the value of $\eta$ was found to be
\begin{eqnarray}
\eta & = & -0.009 \pm 0.033 \pm 0.024. \nonumber
\end{eqnarray}

The nine parameter fit without any assumption of lepton universality gave the
following results:
\begin{eqnarray}
\eta_{\mu} & = & 0.72 \pm 0.32 \pm 0.15, \nonumber \\
\rho_{e} & = & 0.744 \pm 0.036 \pm 0.037, \nonumber \\
\rho_{\mu} & = & 0.999 \pm 0.098 \pm 0.045, \nonumber \\
\xi_{e} & = & 1.01 \pm 0.12 \pm 0.05, \nonumber \\
\xi_{\mu} & = & 1.16 \pm 0.19 \pm 0.06, \nonumber \\
\xi_{e} \delta_{e} & = & 0.85 \pm 0.12 \pm 0.04, \nonumber \\
\xi_{\mu} \delta_{\mu} & = & 0.86 \pm 0.13 \pm 0.04, \nonumber \\
h_{\nu_{\tau}} & = & -0.991 \pm 0.028 \pm 0.011, \nonumber \\
{\cal P}_{\tau} & = & -0.131 \pm 0.012 \pm 0.008. \nonumber
\end{eqnarray}
The parameters are correlated and the correlation matrix is given in
Table~\ref{t:corr9}.
\begin{table}[hbtp]
\begin{center}
 \begin{tabular}{c|r r r r r r r r}
& \multicolumn{1}{c}{$\rho_e$ }
& \multicolumn{1}{c}{$\rho_{\mu}$ }
& \multicolumn{1}{c}{${\cal P}_{\tau}$}
& \multicolumn{1}{c}{$\xi_e$ }
& \multicolumn{1}{c}{$\xi_{\mu}$ }
& \multicolumn{1}{c}{$\xi_e \delta_e$ }
& \multicolumn{1}{c}{$\xi_{\mu} \delta_{\mu}$}
& \multicolumn{1}{c}{$h_{\nu_{\tau}}$ }\\
   \hline
$\eta_{\mu}$      &$-0.102$&$ 0.937$&$-0.065$&$-0.003$&$ 0.678$&$-0.029$&$ 0.423$&$ 0.060$ \\
$\rho_e$          &        &$-0.071$&$ 0.331$&$-0.306$&$ 0.047$&$-0.230$&$ 0.032$&$-0.155$ \\
$\rho_{\mu}$      &        &        &$ 0.062$&$ 0.059$&$ 0.569$&$ 0.012$&$ 0.327$&$-0.006$ \\
${\cal P}_{\tau}$ &        &        &        &$-0.002$&$-0.035$&$-0.110$&$-0.130$&$-0.420$ \\
$\xi_e$           &        &        &        &        &$-0.184$&$ 0.342$&$-0.306$&$ 0.039$ \\
$\xi_{\mu}$       &        &        &        &        &        &$-0.318$&$ 0.415$&$ 0.095$ \\
$\xi_e \delta_e$  &        &        &        &        &        &        &$-0.102$&$ 0.087$ \\
$\xi_{\mu} \delta_{\mu}$ & &        &        &        &        &        &        &$ 0.157$ \\
 \end{tabular}
\caption{The correlation matrix for the nine parameter fit.}
\label{t:corr9}
\end{center}
\end{table}

The values of the Michel parameters for the process $\tau \rightarrow \mu
\overline{\nu}_{\mu} \nu_{\tau}$ are less precisely known than those from
the $\tau \rightarrow \mathrm{e} \overline{\nu}_{e} \nu_{\tau}$ channel. This 
is because for the $\tau \rightarrow \mu \overline{\nu}_{\mu} \nu_{\tau}$
channel one is also measuring the $\eta$ parameter which has all its
sensitivity in this channel.
The $\eta_{\mu}$ and $\rho_{\mu}$ parameters are both 
at the level of $\sim 2 \sigma$ away from
the Standard Model predictions. 
%From the correlation matrix it is clear that 
These two parameters are very highly correlated. In setting
$\eta_{\mu}$ to its Standard Model prediction value of 0 one obtains the 
following results:
\begin{eqnarray}
\eta_{\mu} & = & 0 \, \, \, \mathrm{(fixed)}, \nonumber \\
\rho_{e} & = & 0.755 \pm 0.036 \pm 0.037, \nonumber \\
\rho_{\mu} & = & 0.789 \pm 0.028 \pm 0.012, \nonumber \\
\xi_{e} & = & 1.00 \pm 0.12 \pm 0.05, \nonumber \\
\xi_{\mu} & = & 0.87 \pm 0.11 \pm 0.03, \nonumber \\
\xi_{e} \delta_{e} & = & 0.86 \pm 0.12 \pm 0.04, \nonumber \\
\xi_{\mu} \delta_{\mu} & = & 0.733 \pm 0.094 \pm 0.030, \nonumber \\
h_{\nu_{\tau}} & = & -0.995 \pm 0.028 \pm 0.011, \nonumber \\
{\cal P}_{\tau} & = & -0.129 \pm 0.012 \pm 0.008. \nonumber
\end{eqnarray}

The presented measurements show no deviations from the predictions of 
pure $V\!-\!A$ couplings in $\tau$ decays.

\begin{figure}[hbt]
\centering
  \epsfxsize=10.0cm
  \leavevmode
  \epsfbox[45 146 549 676]{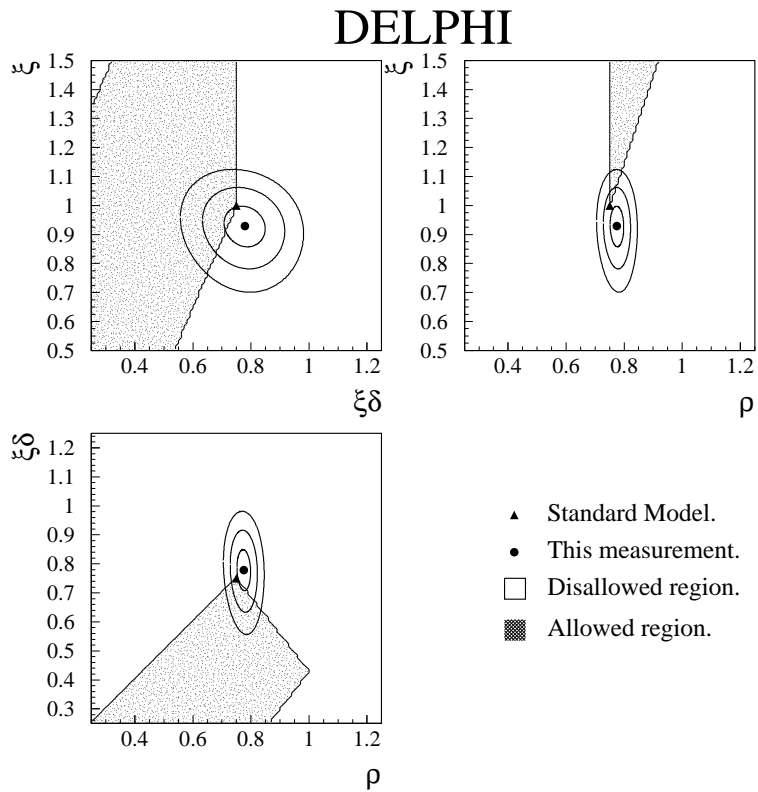}
 \caption{The contours corresponding to $(log {\cal L} + \frac {i^2} {2})$, 
where $i=1,2,3$, for the six parameter fit. In forming the contours the
likelihood function is minimised with respect to the other four parameters 
in the fit. In each plot the allowed region corresponds to the case where 
all the Michel parameters are fixed to their Standard Model values.}
 \label{fig:contours}
\end{figure}

As mentioned in Section~\ref{mich} the Michel parameters are restricted by
boundary conditions. The physically allowed regions for 
various pairs of the parameters $\rho$, $\xi$ and $\xi\delta$ are
shown in Fig.~\ref{fig:contours} along with the experimentally determined
values for these parameters, for the fit assuming lepton universality. 
In forming the contours the likelihood function is minimised with respect to 
the other four parameters in the fit. One can see that the contours enter 
into the disallowed regions due to the finite experimental resolution.
The disallowed regions in Fig.~\ref{fig:contours} 
are in fact dependent on three of the Michel parameters. 
The disallowed regions shown are presented with the Michel parameters set at 
their Standard Model values for simplicity. These regions will therefore 
move around to encompass more of the fitted contours if the Michel parameters 
are set at their measured values. 
%The contours can also enter the disallowed
%regions due to inherent effects of the smearing of the fit variables due to 
%the imperfect ability of the detector to reconstruct these variables and 
%also due to the fact that the data are binned in the Michel parameter fit.

\boldmath
\section{Extraction of the coupling $\kappa^W_\tau$}
\unboldmath  The spectra  of the  $\tau$ decay  products were  used to
extract  the   parameter  $\kappa^W_\tau$.  To   estimate  the  theory
prediction  of the spectra  distortion in  the case  of $\kappa^W_\tau
\neq 0$  the Standard Model simulated  data were used  with the events
re-weighted  in the  following way.  For the  generated values  of the
$\tau$   helicity   and  the   final   lepton   momentum,  the   value
$d\Gamma/dx_l(x_l,\kappa^W_\tau)$   was    calculated   according   to
Eqns.~\ref{eqn:width}      and     \ref{eqn:polys}.      The     ratio
$\frac{d\Gamma/dx_l(x_l,\kappa^W_\tau)}{d\Gamma/dx_l(x_l,0)}$ was then
used  as  an event  weight  to  produce  the simulated  spectrum  with
non-zero tensor coupling. In the case of the $\tau$ multipionic decays
Eqn.~\ref{eqn:semilep}  was  used  to  generate  event  weights.   The
radiative corrections to $\tau$ production  and $\tau$ decay were taken into
account by the KORALZ~4.0 and TAUOLA~2.5 programs and the variable $x_l$
was  defined using  the lepton  energy  after all  radiation.  It  was
assumed  the effect  of the  tensor coupling  was small  and  that the
effects of radiative corrections  to the tensor coupling contributions
could be neglected.

The value of the tensor coupling parameter was then extracted from a
log likelihood fit of the simulated spectra to the real data, with
$\kappa^W_\tau$ as a fit parameter.
One-dimensional spectra of $x_l$ were used in the case of leptonic $\tau$ decays
and the two-dimensional spectra of $(\cos{\theta^*}, \cos{\psi})$ for
semi-leptonic decays. To increase the sensitivity of the semi-leptonic 
channel further, the region of reconstructed invariant mass between 
$0.3$ and $1.7$ $\mathrm{GeV/c^2}$ was divided
into five bins and the fit was performed in each bin simultaneously. This
reduced the statistical error of the fit by about 10\%. The region
of invariant mass below 0.3 $\mathrm{GeV/c^2}$ was not used because it was 
dominated by $\tau \rightarrow \pi \nu_{\tau}$ decays which have no sensitivity
to the tensor coupling.

\begin{figure}[hbt]
  \begin{center}
     \leavevmode
\put(140,270){\makebox(0,0){\Large \bf DELPHI}}
\epsfig{figure=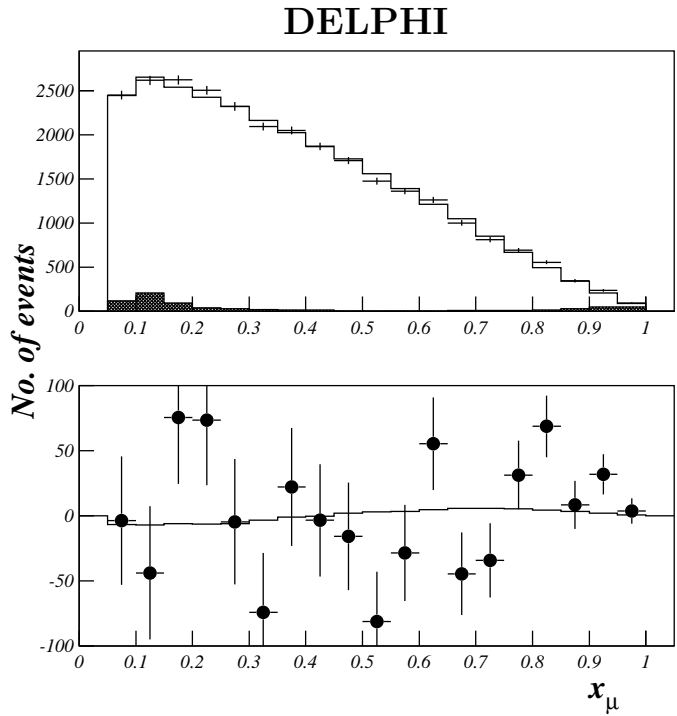,width=10cm,bbllx=80,bblly=200,bburx=510,bbury=630} 
\caption{
An illustration of the fit to the tensor coupling parameter using the decay
$\tau \rightarrow \mu \overline{\nu}_{\mu} \nu_{\tau}$. Upper plot: spectra
of the normalised muon momentum for data (points with error bars), background
(black) and the best fit simulated data (solid line). Lower plot: the
difference between the measured spectrum and the Standard Model prediction
(points with error bars); the solid line shows the difference between the best
fit simulation and the Standard Model simulation.}
\label{fig:tensor}
  \end{center}
\end{figure}
The illustration for
the channel $\tau \rightarrow \mu \overline{\nu}_{\mu} \nu_{\tau}$ is given
in Fig.~\ref{fig:tensor} 
which shows the difference between the real data and the Standard Model simulated
data prediction. Also shown is the difference between the best fit simulated data 
and the Standard Model simulated data. 
The sample used for this channel was an 
equivalent data set to that used for the Michel parameter analysis together 
with $\tau \rightarrow \mu \overline{\nu}_{\mu} \nu_{\tau}$ decays detected 
in the end-cap region of the detector. The selection follows that described 
in~\cite{delphipolarisation}. 

The systematic uncertainty for the muon channel received contributions
from the limited simulation statistics, the dimuon background level and
the calibration of the momentum scale of the charged particle track reconstruction.
For the $ \mathrm{e} \overline{\nu}_{e} \nu_{\tau}$ final state
the main contributions to the systematic uncertainty are the limited
simulation statistics, the level of the Bhabha background and
the calibration of the electron energy estimator~$p_{el}$.
For the hadronic selection the main systematic is the calibration
of the neutral electromagnetic shower energy scale. Other significant
contributions arise from the efficiency of photon detection in the HPC and
the knowledge of the resolution on the fit quantities $\theta^*$ and $\psi$.
The calibration of the momentum scale for the charged hadron also gives
a contribution.
Other sources of systematic uncertainty common to some or all channels are
the modelling of the momentum resolution and the electromagnetic energy
resolution, and the uncertainty on the $\tau$
branching ratios. 
%The systematic uncertainty received
%contributions from the limited simulation statistics, the background level 
%uncertainty, the variation of spectrum binning, the momentum scale 
%uncertainty for charged particles and photons, misidentification of 
%$\tau \rightarrow \rho \nu_{\tau}$ candidates as $\tau \rightarrow \pi
%\nu_{\tau}$ and the uncertainties in the world average values of the $\tau$
%branching ratios.

The results of the fits for different decay channels were the following:
\begin{eqnarray}
\tau \rightarrow \mathrm{e} \overline{\nu}_{e} \nu_{\tau} & : &
\kappa^W_\tau = +0.162 \pm 0.078 \pm 0.030, \nonumber \\
\tau \rightarrow \mu \overline{\nu}_{\mu} \nu_{\tau} & : &
\kappa^W_\tau = -0.043 \pm 0.057 \pm 0.032, \nonumber \\
\tau \rightarrow h (n \pi^0) \nu_{\tau} & : &
\kappa^W_\tau = -0.122 \pm 0.059 \pm 0.025. \nonumber
\end{eqnarray}
\noindent
Combining these taking into account correlations gave
$$\kappa^W_\tau  =  -0.029 \pm  0.036  \pm  0.018,$$  where the  first
uncertainty is statistical and  the second is systematic.  The $\chi^2
/  N_{dof}$   for  this  combination  is  7.6/2   corresponding  to  a
probability  of having  a  worse  $\chi^2 /  N_{dof}$  of 2.3\%.   The
measured  value  of  the  anomalous tensor  coupling,  $\kappa^W_\tau$
corresponds to a 
$90\%$ allowed interval of $-0.096 < \kappa^W_\tau < 0.037$.
%$95 \%$ allowed interval of $-0.108 < \kappa^W_\tau < 0.050$.

\section{Interpretation of the results}
\label{sec:interp}

The results of this analysis can be used 
and interpreted in a number of different ways. It is interesting now to use 
the measured values of the Michel parameters to explore these avenues and 
investigate the possible existence of new physics beyond the Standard 
Model. 

The measured values of the parameters $\xi$ and $\xi \delta$ 
were used to estimate the 
probability $P_{R}^{\tau}$ of a right handed tau decaying into a lepton of 
either handedness. 
Following the technique outlined in~\cite{feldman}, and
only allowing the value of $P^{\tau}_R$ to be between 0 and 1, the 
corresponding upper limit on this quantity was found to be:
$$ P^\tau_R < 0.081 
{\rm \, \, \, at \, \, 90\% \, \, C.L.}$$
Using the constraints described in Section~\ref{mich} one can set limits 
on the coupling constants by forming positive definite 
expressions from the measured parameters. 
The best limits on $g^S_{LR}$ and $g^T_{LR}$ are 
derived from~Eqn.~\ref{eqn:ptaur} using the limit on $P_{R}^{\tau}$.
Eqn.~\ref{eqn:ineq4} constrains $g^V_{LR}$, while the best constraints on
$g^S_{RR}$ and $g^V_{RR}$ are derived from Eqn.~\ref{eqn:ineq3}.
The best limits on the $g^V_{RL}$, $g^S_{RL}$ and $g^T_{RL}$
are obtained from~Eqn.~\ref{eqn:ineq1}. 
Absorbing the freedom given by the unknown overall phase,
the coupling $g^V_{LL}$ is taken to be real and positive.
Only the coupling $g^S_{LL}$ cannot
be constrained as one cannot distinguish between $g^S_{LL}$ and the Standard 
Model coupling $g^V_{LL}$. It could be constrained by measuring the
cross-section for inverse $\tau$ decay~\cite{fetscher}.  
The overall normalisation factor $A$ in 
Eqns.~\ref{eqn:decaywidth}~and~\ref{eqn:normalisation} was fixed
to a value~of~16.

\begin{figure}[hbt]
\centering
  \leavevmode
\epsfig{figure=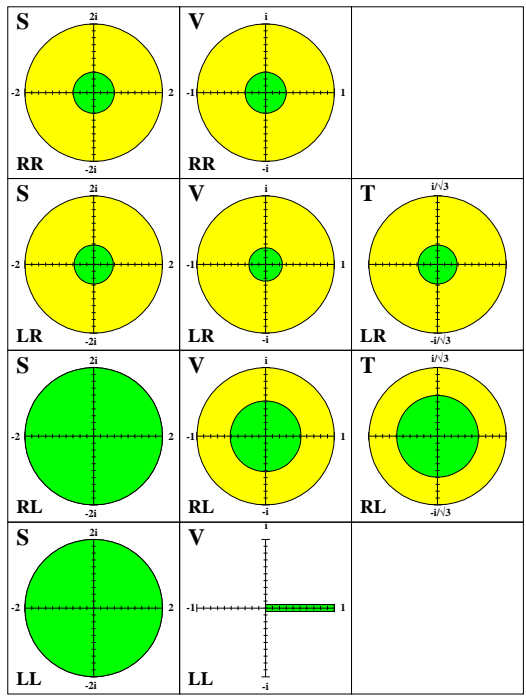,width=8cm} 
%  \epsfbox[93 143 502 736]{p1.ps}
%  \epsfbox[]{p1new.eps}
\caption{Limits (at 90\% CL) on the coupling constants from the Michel 
parameter fit assuming lepton universality. The dark shaded areas are the allowed
regions.}
 \label{fig:lim_uni}
\end{figure}

The 90\% confidence level upper limits 
on the coupling constants, derived as  described in~\cite{feldman},
are given in Table~\ref{tab:couplings} for the fits with and without lepton 
universality. The parameters $g^S_{RL}$ and $g^T_{RL}$
are coupled together in ~Eqn.~\ref{eqn:ineq1}.
The limits obtained for $g^T_{RL}$ assume $g^S_{RL}=0$; if this 
condition is relaxed, the limits obtained for $g^T_{RL}$ are poorer than the
normalisation constraint.
The results of the fit assuming universality are
also illustrated pictorially in Fig.~\ref{fig:lim_uni}.
\begin{table}[hbtp]
\begin{center}
 \begin{tabular}{c|ccc|c}
   \hline
coupling  &  e-$\mu$ & e& $\mu$ &maximum\\
   \hline
$g^S_{RR}$  & ~0.598~ &  ~0.765~  & ~0.999~    &  2  \\
$g^S_{LR}$  & 0.568 &  0.805  & 0.791    &  2 \\
$g^S_{RL}$  & 2.000   &  2.000    & 2.000    &  2 \\
$g^S_{LL}$  &  -    &   -     &  -       &  2 \\
   \hline                                  
$g^V_{RR}$  & 0.299 &  0.382  & 0.499    &  1 \\
$g^V_{LR}$  & 0.243 &  0.397  & 0.302    &  1 \\
$g^V_{RL}$  & 0.515 &  0.564  & 0.422    &  1 \\
$g^V_{LL}$  &  -    &   -     &  -       &  1 \\
   \hline                                  
$g^T_{LR}$  & 0.164 &  0.232  & 0.228    &  $1/\sqrt{3}$ \\
$g^T_{RL}$  & 0.343 &  0.387  & 0.281    &  $1/\sqrt{3}$ \\
%$g^T_{RL}$  & 0.577 &  0.577  & 0.577   &  $1/\sqrt{3}$ \\
%$g^T_{RL}$  & 0.677 &  0.720  & 0.614   &  $1/\sqrt{3}$ \\
   \hline
 \end{tabular}
\caption{90\% C.L. upper limits on the magnitudes of the complex coupling constants.
The 2nd column contains the results assuming e-$\mu$ universality. The 3rd and 4th
columns display the results for the electronic and muonic decay modes respectively.
The fifth column shows the maximum physically allowed value for the parameter.}
\label{tab:couplings}
\end{center}
\end{table}

In its minimal version, the Higgs mechanism is implemented by adding
only one doublet of complex scalar fields resulting in one additional
physical scalar state, electrically neutral, commonly referred to as
the standard Higgs boson. One can postulate extensions to this by
adding, for example, one more doublet of complex scalar fields, which
leads automatically to five physical states (three neutral and a
pair of charged Higgs bosons), after the spontaneous
breaking of the $\mathrm{SU(2)_L\times U(1)_Y}$ symmetry to give mass
to the $W^\pm$ and $Z$ gauge bosons. These charged Higgs bosons contribute to
the $\tau\to l \bar{\nu}_l {\nu}_{\tau}$, $l$=e,$\mu$ decay
through a scalar coupling, given at Born level by
\begin{equation}
g_{l}^{S} = - \frac {m_l m_{\tau} 
      \mathrm{tan}^2 \beta}{ m^2_{H^{\pm}}},
\end{equation}
for negligible neutrino
masses~\cite{stahl,habkan,mcwilli}. $\mathrm{Tan \beta}$ is the ratio of the
vacuum expectation values of the two Higgs doublets and $m_{H^{\pm}}$
is the mass of the charged Higgs boson. For left-handed neutrinos the
couplings are of the type $g^S_{RR}$, and the Michel parameters
$\eta_l$ and $\xi_l$, (with $l$=e,$\mu$),
can be written in terms of $g^S_{RR,l}$:
\begin{eqnarray}
\eta_l & = & - \frac {g^S_{RR,l} / 2}
            {1 + (g^S_{RR,l} / 2)^2}; \label{eqn:higgs1} \\  
\xi_l & = & \frac {1 - (g^S_{RR,l} / 2)^2} {1 + (g^S_{RR,l} / 2)^2},\label{eqn:higgs2}
\end{eqnarray}
while $\rho=\delta=\textstyle\frac{3}{4}$. With these relations $m_{H^{\pm}}$
can be extracted using the measured values of the Michel parameters.
Using the 
results presented in this paper one obtains a lower limit on the mass of 
the charged Higgs of \footnote{the limits presented for the two doublet and
left-right symmetric models correspond to the value at which the likelihood 
has dropped to give the corresponding measure of confidence.}
$$m_{H^{\pm}} > 1.17 \times \tan \beta 
{\rm \, \, \, GeV/c^2 \, \, \, at \, \, 90\% \, \, C.L.}$$
%where $\tan \beta$ is the ratio of the vacuum expectation values of the
%neutral components of the two charged Higgs fields. 
This limit is not competitive with those from direct searches, 
unless $\tan \beta$ has an unexpectedly large value.

\begin{figure}[hbt]
\centering
  \epsfxsize=10.0cm
  \leavevmode
  \epsfbox[0 0 560 560]{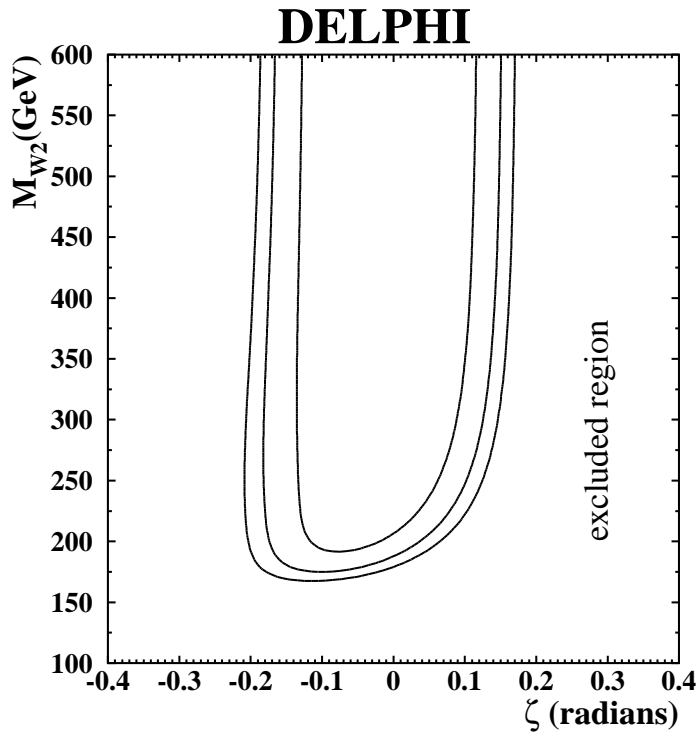}
 \caption{Contours corresponding to 68\%, 95\% and 99\% confidence levels on 
the~$\zeta$ versus~$m_{W_2}$ plane for the fit to the left-right symmetric model.}
 \label{fig:zetamass2d}
\end{figure}

Another extension to the Standard Model which can be related to the
Michel parameters involves the postulate that parity violation is
caused by spontaneous symmetry breaking. A question which arises in
the Standard Model is why the doublets are left handed and the
singlets right-handed. Left-right symmetry implies that the
Lagrangian is both charge and parity invariant before spontaneous
symmetry breaking and that CP violation arises due to the
non-invariance of the vacuum. These left-right symmetric models assume
the existence of a second pair of $W$ bosons, and the weak eigenstates $W_{L,R}$
are mixtures of the mass eigenstates $W_{1,2}$~\cite{polak1,polak2}. 
One can introduce the mass ratio\footnote{this mass ratio is 
commonly referred to as $\beta$ in the literature. In order to avoid 
confusion with the parameter introduced above, $\tan \beta$, the name has
been changed to $\alpha$} $\alpha$ of the mass eigenstates,
\begin{equation}
\alpha = m^2_{W_1} / m^2_{W_2},
\end{equation}
and the mixing angle $\zeta$ between the weak and mass eigenstates. 
%as
%\begin{equation}
%m^2_{W_{1,2}} = \frac 1 2 \biggl(
%m^2_{W_L} + m^2_{W_R} \pm 
%\frac {m^2_{W_L} - m^2_{W_R}} {\cos 2 \zeta} \biggr) .
%\end{equation}
The Michel parameters $\rho$ and $\xi$ take the form
\begin{eqnarray}
\rho & = & \frac 3 4 \cos^4 \zeta \biggl( 
1 + \tan^4 \zeta + \frac {4 \alpha} {1 + \alpha^2} \tan^2 \zeta \biggr) ,
\label{eqn:lr1} \\
\xi & = & \cos^2 \zeta (1 - \tan^2 \zeta)
\frac {1 - \alpha^2} {1 + \alpha^2}, \label{eqn:lr2}
\end{eqnarray}
while $\eta=0$ and $\delta=\textstyle\frac{3}{4}$.
The  $\nu_\tau$ polarisation parameter in $\tau$
hadronic decays takes the form~\cite{stahlprivate}
\begin{equation}
h_{\nu_\tau} = -1+2\zeta^2+\alpha(4\zeta-8\zeta^2)
               +\alpha^2(2-8\zeta+12\zeta^2).
\end{equation}
Using these relations, the measured value of $m_{W_1}$ and the
measured values of the Michel parameters, one can place limits 
on~$m_{W_2}$ and~$\zeta$, with the caveat that the 
right-handed neutrino must be light enough
compared with the $\tau$ to be produced 
without kinematical suppression.
Taking the measured 
values of the Michel parameters and 
$m_{W_1} = (80.41 \pm 0.10)~\mathrm{GeV/c^2}$~\cite{pdg98} 
gives the following limits on $m_{W_2}$ and 
the mixing angle $\zeta$:
$$m_{W_2} > 189
{\rm \, \, \, GeV/c^2 \, \, \, at \, \, 90\% \, \, C.L.};$$
$$ -0.141 < \zeta < 0.125~  {\mathrm rad}
{\rm \, \, \, at \, \, 90\% \, \, C.L.}$$
The 68\%, 95\% and 99\% confidence level
contours on the $\zeta$-$m_{W_2}$
plane are shown in Fig.~\ref{fig:zetamass2d}.
The $\chi^2$ function exhibits a slight minimum 
at~$m_{W_2}=290$~GeV/$c^2$,~$\zeta=-0.01$. 
%The $\chi^2$ at this point is 1.52 for 4 d.o.f. 
The distribution then exhibits an allowed region extending 
to infinite $m_{W_2}$ where the change in~$\chi^2$, for~$\zeta=0$, is~0.13 compared with
the minimum. For the case of no mixing $\zeta=0$, and the lower limit on the $W_2$  mass
becomes
$$m_{W_2} > 204
{\rm \, \, \, GeV/c^2 \, \, \, at \, \, 90\% \, \, C.L.}.$$

\section{Conclusions}

A precise measurement of the Michel parameters and the
$\nu_{\tau}$ helicity has been presented, together with limits on
the anomalous tensor coupling. 

A simultaneous
fit to the Michel parameters and the $\nu_\tau$ helicity
assuming e-$\mu$ universality and using the DELPHI exclusive
leptonic branching ratio measurements~\cite{bjarne}
gave the following results:
\begin{eqnarray}
\eta & = & -0.005 \pm 0.036 \pm 0.037, \nonumber \\
\rho & = & 0.775 \pm 0.023 \pm 0.020, \nonumber \\
\xi & = & 0.929 \pm 0.070 \pm 0.030, \nonumber \\
\xi \delta & = & 0.779 \pm 0.070 \pm 0.028, \nonumber \\
h_{\nu_{\tau}} & = & -0.997 \pm 0.027 \pm 0.011. \nonumber
\end{eqnarray}

A fit to the Michel parameters and the $\nu_\tau$ helicity
not assuming universality
gave the following results:
\begin{eqnarray}
\eta_{\mu} & = & 0.72 \pm 0.32 \pm 0.15, \nonumber \\
\rho_{e} & = & 0.744 \pm 0.036 \pm 0.037, \nonumber \\
\rho_{\mu} & = & 0.999 \pm 0.098 \pm 0.045, \nonumber \\
\xi_{e} & = & 1.01 \pm 0.12 \pm 0.05, \nonumber \\
\xi_{\mu} & = & 1.16 \pm 0.19 \pm 0.06, \nonumber \\
\xi_{e} \delta_{e} & = & 0.85 \pm 0.12 \pm 0.04, \nonumber \\
\xi_{\mu} \delta_{\mu} & = & 0.86 \pm 0.13 \pm 0.04, \nonumber \\
h_{\nu_{\tau}} & = & -0.991 \pm 0.028 \pm 0.011. \nonumber
\end{eqnarray}
In both fits the average $\tau$ polarisation was left as a  free
parameter.

The world averages for the $\nu_{\tau}$ helicity and for
the Michel parameters $\rho$, $\xi$ and $\xi\delta$,
both with and without the assumption of lepton universality,
are dominated by the~results~\cite{cleo3} from the CLEO experiment.
The results presented here are 
consistent with the world averages and have a precision between~1.5 and~2.5 times
poorer than the CLEO measurements
for these parameters but are of a similar or higher precision than other 
measurements, taken at LEP~\cite{opal,aleph,l31,l32} or by ARGUS~\cite{argus6}. 

Fixing $h_{\nu_{\tau}}$ to the standard model value of $-1$
and constraining ${\cal P}_{\tau}$ from other measurements of the effective weak mixing
angle would give a reduction in the quoted errors on the Michel parameters
due to the correlations present in the simultaneous fits to all
the parameters.
 
The measurement of $\eta$ presented here is the most precise recorded to date,
due to the combination of the precise leptonic branching ratio measurements
and the measurements of the spectra.

A measurement of the tensor coupling $\kappa_\tau^W$
has been performed for the first time in $\tau$ decays, yielding the result
$$\kappa^W_\tau = -0.029 \pm 0.036 \pm 0.018,$$
consistent with zero.

The presented results are consistent with the Standard Model and
limits have been placed on the magnitudes of the 
complex coupling constants. The $V\!-\!A$ assumption is
however still not fully verified. 
Future results from $B$ factories, 
complemented by a measurement of inverse $\tau$ decay, will
allow a full determination of the Lorentz structure of the~$\tau$.

%\newpage
%         Modified on 04-06-1999 by dimartino
%-------------------------------------------------------------------
\subsection*{Acknowledgements}
\vskip 3 mm
\noindent The authors gratefully acknowledge Achim Stahl for 
his help in interpreting the results presented in this paper.\\
 We are also greatly indebted to our technical 
collaborators, to the members of the CERN-SL Division for the excellent 
performance of the LEP collider, and to the funding agencies for their
support in building and operating the DELPHI detector.\\
We acknowledge in particular the support of \\
Austrian Federal Ministry of Science and Traffics, GZ 616.364/2-III/2a/98, \\
FNRS--FWO, Belgium,  \\
FINEP, CNPq, CAPES, FUJB and FAPERJ, Brazil, \\
Czech Ministry of Industry and Trade, GA CR 202/96/0450 and GA AVCR A1010521,\\
Danish Natural Research Council, \\
Commission of the European Communities (DG XII), \\
Direction des Sciences de la Mati$\grave{\mbox{\rm e}}$re, CEA, France, \\
Bundesministerium f$\ddot{\mbox{\rm u}}$r Bildung, Wissenschaft, Forschung 
und Technologie, Germany,\\
General Secretariat for Research and Technology, Greece, \\
National Science Foundation (NWO) and Foundation for Research on Matter (FOM),
The Netherlands, \\
Norwegian Research Council,  \\
State Committee for Scientific Research, Poland, 2P03B06015, 2P03B1116 and
SPUB/P03/178/98, \\
JNICT--Junta Nacional de Investiga\c{c}\~{a}o Cient\'{\i}fica 
e Tecnol$\acute{\mbox{\rm o}}$gica, Portugal, \\
Vedecka grantova agentura MS SR, Slovakia, Nr. 95/5195/134, \\
Ministry of Science and Technology of the Republic of Slovenia, \\
CICYT, Spain, AEN96--1661 and AEN96-1681,  \\
The Swedish Natural Science Research Council,      \\
Particle Physics and Astronomy Research Council, UK, \\
Department of Energy, USA, DE--FG02--94ER40817. \\
%=========================================================================%

\end{document}